\documentclass[journal,draftcls,onecolumn,letterpaper,twoside,11pt]{IEEEtran}

\usepackage[cmex10]{amsmath}
\usepackage{latexsym,amssymb,cite,bbm,epsf,amsthm,subfigure,algorithm,algorithmic}
\usepackage{graphicx}
\usepackage{enumerate}
\usepackage{mathrsfs}
\usepackage{epstopdf}
\usepackage{bbm}

\newcommand{\bc}{\begin{center}}
\newcommand{\ec}{\end{center}}
\newcommand{\be}{\begin{equation}}
\newcommand{\ee}{\end{equation}}
\newcommand{\ba}{\begin{array}}
\newcommand{\ea}{\end{array}}
\newcommand{\bea}{\begin{eqnarray}}
\newcommand{\eea}{\end{eqnarray}}
\newcommand{\bal}{\begin{align}}
\newcommand{\eal}{\end{align}}
\newcommand{\ei}{\end{itemize}}
\newcommand{\bi}{\begin{itemize}}
\newcommand{\bfi}{\begin{figure}}
\newcommand{\efi}{\end{figure}}
\newcommand{\MB}{\left[\begin{array}}
\newcommand{\ME}{\end{array}\right]}
\newcommand{\nn}{\nonumber}

\newtheorem{thm}{Theorem}
\newtheorem{dfn}{Definition}
\newtheorem{cor}{Corollary}
\newtheorem{lem}{Lemma}

\newcommand{\Exp}{\mathsf{E}}
\newcommand{\Pro}{\mathsf{P}}
\newcommand{\Hyp}{\mathsf{H}}
\newcommand{\I}{\mathsf{I}}

\newcommand{\cS}{\mathcal{S}}
\newcommand{\cT}{\mathcal{T}}
\newcommand{\cH}{\mathcal{H}}
\newcommand{\cM}{\mathcal{M}}
\newcommand{\cN}{\mathcal{N}}

\newcommand{\ind}[1]{\mathbbm{1}_{\{#1\}}}   

\newcommand{\ignore}[1]{{}}

\begin{document}

\title{Cooperative Sequential Spectrum Sensing Based on Level-triggered Sampling}

\author{Yasin Yilmaz\IEEEauthorrefmark{2}\footnote{\IEEEauthorrefmark{2}Electrical Engineering Department, Columbia University, New York, NY 10027.},\;\; George V. Moustakides\IEEEauthorrefmark{3}\footnote{\IEEEauthorrefmark{3}Dept. of Electrical \& Computer Engineering, University of Patras, 26500 Rion, Greece.},\;\; and \, Xiaodong Wang\IEEEauthorrefmark{2}}

\maketitle

\begin{abstract}
We propose a new framework for cooperative spectrum sensing in cognitive radio networks, that
is based on a novel class of non-uniform samplers, called the event-triggered samplers, and sequential
detection. In the proposed scheme, each secondary user computes its local sensing decision statistic based
on its own channel output; and whenever such decision statistic crosses certain predefined threshold
values, the secondary user will send one (or several) bit of information to the fusion center. The fusion
center asynchronously receives the bits from different secondary users and updates the
global sensing decision statistic to perform a sequential probability ratio test (SPRT), to reach
a sensing decision. We provide an asymptotic analysis for the above scheme, and under different
conditions, we compare it against the cooperative sensing scheme that is based
on traditional uniform sampling and sequential detection.
Simulation results show that the proposed scheme, using even $1$ bit, can outperform its uniform sampling counterpart that uses infinite number of bits under changing target error probabilities, SNR values, and number of SUs.
\end{abstract}

\section{Introduction}\label{sec:intro}

Spectrum sensing is one of the most important functionalities in
a cognitive radio system \cite{Yucek_09}, by which the secondary users
(SU) decide whether or not the spectrum is being used by the primary
users. Various spectrum sensing methods have been developed based on
exploiting different features of the primary user's signal \cite{Ma_09}.
On the other hand, cooperative sensing,
where multiple secondary users monitor the
spectrum band of interest simultaneously and cooperate to make a
sensing decision, is an effective way to achieve fast and reliable
spectrum sensing \cite{Ganesan_07,Sun_07,veeravalli_07,Quan_08,Ma_08}.

In cooperative sensing, each secondary user collects its own local
channel statistic, and sends it to a fusion center (FC), which then
combines all local statistics received from the secondary users
to make a final sensing decision. The decision mechanism at the FC
can be either sequential or fixed sample size.
In other words, the FC can either try to make a decision every time it receives new
information or it can wait to collect a specific number of samples and then
make a final decision using them.
It is known that sequential methods are
much more effective in minimizing the decision delay than their fixed sample
size counterparts. In particular, the
sequential probability ratio test  (SPRT)
 is the dual of the fixed sample size Neyman-Pearson test,
and it is optimal among all sequential tests in terms of minimizing the
average sample number (decision delay) for i.i.d. observations \cite{Wald_47,Wald_48}.
Sequential approaches to spectrum sensing have been proposed in a number of recent works
\cite{Poor_09,Sharma_11,Kim_10,Kundargi_10,Shei_08,Xin_09}.

\ignore{
All existing works on cooperative and sequential sensing assume that
the local and global decision statistics are obtained based on the conventional
uniform sampling of the underlying signals. }

The majority of existing works on cooperative and sequential sensing assume that
the SUs synchronously communicate to the FC. This implies the existence of a global clock according to which SUs sample their local test statistics using conventional uniform sampling.
There are a few works allowing for asynchrony among SUs (e.g., \cite{Kundargi_10,Shei_08}), but none of them provides an analytical discussion on the optimality or the efficiency of the proposed schemes.
In this paper, we develop a new
framework for cooperative sensing   based on
  a class of non-uniform samplers called the \emph{event-triggered samplers},
in which the sampling times are
determined in a dynamic way by the signal to be sampled.
Such a sampling scheme naturally outputs low-rate information (e.g., 1 bit per sample)
without performing any quantization, and permits asynchronous communication between
the SUs and the FC \cite{Fellouris11}. Both features are ideally suited for  cooperative sensing in cognitive
radio systems since the control channel for transmitting local statistics has a low
bandwidth and it is difficult to maintain synchrony among the SUs. Moreover, we will show that by properly designing
the operations at the SUs and FC, the cooperative sensing scheme based on
event-triggered sampling can outperform the one based on the conventional uniform sampling.

The remainder of the paper is organized as follows. In Section II we
describe the  cooperative spectrum sensing problem, in both centralized
and decentralized form and we outline three spectrum
detectors. In Section III we introduce the decentralized
spectrum sensing approach based on event-triggered sampling and discuss the operations at both the SUs and the FC.
In Section IV we perform a comprehensive asymptotic performance
analysis on the proposed spectrum sensing method, and the
one based on conventional uniform sampling. Simulation results
are provided in Section V. Finally, Section VI concludes the paper. 

\section{Problem Formulation and Background}
\label{sec:ProbState}

\subsection{Spectrum Sensing via SPRT}
\label{sec:ProbState1}

Consider a cognitive radio network where there are $K$ secondary
users performing spectrum sensing and dynamic spectrum access.
Let $\{y^k_{t}\},~t \in {\mathbb N}$,   be the Nyquist-rate sampled
discrete-time signal   observed by the $k$th SU, which processes it
and transmits some form of local information  to a fusion
center.
Using the information received at the fusion center from the $K$
SUs, we are interested in deciding between two hypotheses,
$\Hyp_0$ and $\Hyp_1$, for the SU signals: i.e.,
whether the primary user (PU) is present ($\Hyp_1$) or not
($\Hyp_0$). Specifically,
every time the fusion center receives new information, it performs a
test and either 1) stops accepting more data and decides between the
two hypotheses;  or 2) postpones its decision until a new data
sample arrives from the SUs. When the fusion center stops and
selects between the two hypotheses, the whole process is terminated.

Note that the decision mechanism utilizes the received data sequentially as they arrive at the fusion center. This
type of test is called sequential as opposed to the conventional fixed sample size test in which one waits until a
specific number of samples has been accumulated and then uses them to make the final hypothesis selection.
Since the pioneering work of Wald \cite{Wald_47}, it has been observed that sequential methods require, on average, approximately four times \cite[Page 109]{Poor} less samples (for Gaussian signals) to reach a decision than their fixed sample size counterparts, for the same level of confidence. Consequently, whenever possible, it is always preferable to use sequential over fixed sample size approaches.

Assuming independence across the signals observed by different SUs, we can cast our problem of interest as the following \textit{binary} hypothesis testing problem
\begin{align}
\begin{split}
\Hyp_0:~~& \{y^1_1,\ldots,y^1_t\} \sim f^1_0;~\{y^2_1,\ldots,y^2_t\} \sim f^2_0;\ldots;\{y^K_1,\ldots,y^K_t\} \sim f^K_0 \\
\Hyp_1:~~& \{y^1_1,\ldots,y^1_t\} \sim f^1_1;~\{y^2_1,\ldots,y^2_t\} \sim f^2_1;\ldots;\{y^K_1,\ldots,y^K_t\} \sim f^K_1,
\end{split}
\end{align}
where $\sim$ denotes ``distributed according to'' and $f^k_0$ and $f^k_1$ are the joint probability density functions of the received
signal by the $k$th SU, under $\Hyp_0$ and $\Hyp_1$ respectively. Since we assume independence across different SUs the log-likelihood ratio (LLR) $L_t$ of all the signals received up to time $t$, which is a sufficient statistic for our problem, can be split as
\begin{equation}
L_t=\sum_{k=1}^K L_t^k
\label{eq:1}
\end{equation}
where $L_t^k$ represents the \textit{local} LLR of the signal received by the $k$th SU, namely
\begin{equation}
L_t^k\triangleq\log\frac{f^k_1(y_1^k,\ldots,y_t^k)}{f^k_0(y_1^k,\ldots,y_t^k)}.
\end{equation}
Hence each SU can compute its own LLR based on its corresponding observed signal, and send it to the
fusion center which collects them and computes the global cumulative LLR $L_t$ using \eqref{eq:1}.
Note that the local LLRs can be obtained recursively. That is, at each time $t$, the
new observation $y^k_{t}$ gives rise to an LLR increment $l^k_t$, and the local cumulative LLR can then
be updated as
\begin{equation}
\label{eq:recursive}
L^k_t = L^k_{t-1} + l^k_t=\sum_{n=1}^tl^k_n,
\end{equation}
where
\begin{equation}
l^k_t\triangleq\log\frac{f^k_1(y^k_t|y^k_1,\ldots,y^k_{t-1})}{f^k_0(y^k_t|y^k_1,\ldots,y^k_{t-1})},
\label{eq:sampleLLR}
\end{equation}
and $f^k_i(y^k_t|y^k_1,\ldots,y^k_{t-1})$ denotes the conditional pdf of $y^k_t$ given the past (local) signal samples under hypothesis $\Hyp_i$. Of course when the samples of the received signal in each SU are also i.i.d., that is, we have independence across time, then the previous expression simplifies considerably and we can write
$l^k_t=\log\frac{f^k_1(y^k_t)}{f^k_0(y^k_t)}$ where now $f^k_i$ represents the pdf of a single sample in the $k$th SU under hypothesis $\Hyp_i$.

As we mentioned, the fusion center collects the local LLRs and at each time instant $t$ is faced with a decision, namely
to wait for more data to come, or to stop receiving more data and select one of the two hypotheses $\Hyp_0,\Hyp_1$.
In other words the sequential test consists of a pair $(\cT,\delta_{\cT})$ where $\cT$ is a \textit{stopping time} that decides when to stop (receiving more data) and $\delta_{\cT}$ a selection rule that selects one of the two hypotheses based on the information available up to the time of stopping $\cT$.

Of course the goal is to make a decision as soon as possible which means that we would like to minimize the delay $\cT$, on average, that is,
\begin{equation}
\min_{\cT,\delta_{\cT}}\Exp_0[\cT],~~\text{and/or}~~\min_{\cT,\delta_{\cT}}\Exp_1[\cT].
\label{eq:optim}
\end{equation}
At the same time we would also like to assure the satisfactory performance of the decision mechanism through suitable constraints on the Type-I and Type-II error probabilities, namely
\begin{equation}
    \Pro_0(\delta_{\cT}=1) \le \alpha~\text{and}~\Pro_1(\delta_{\cT}=0)\le\beta,\label{eq:constr}
\end{equation}
where $\Pro_i(\cdot),\Exp_i[\cdot],~i=0,1$ denote probability and the corresponding expectation under hypothesis $i$. Levels $\alpha,\beta\in(0,1)$ are parameters specified by the designer.

Actually, minimizing in \eqref{eq:optim} each $\Exp_i[\cT],~i=0,1$ over the pairs $(\cT,\delta_{\cT})$ that satisfy the two constraints in \eqref{eq:constr}, defines \textit{two} separate constrained optimization problems. However, Wald first suggested \cite{Wald_47} and then proved \cite{Wald_48} that the Sequential Probability Ratio Test (SPRT) \textit{solves both problems simultaneously}. SPRT consists of the pair $(\cS,\delta_{\cS})$ which is defined as follows
    \begin{align}
    \label{eq:SPRT}
    \cS=\inf\left\{t>0: L_t\not\in(-B,A)\right\}, ~~
  \delta_{\cS} &= \left\{ \begin{array}{ll}
        1, & \mbox{if $L_{\cS}\geq A$}, \\
        0, & \mbox{if $L_{\cS}\leq -B$}.
        \end{array} \right.
    \end{align}
In other words, at every time instant $t$, we compare the running LLR $L_t$ with two thresholds $-B,A$ where $A,B>0$. As long as $L_t$ stays within the interval $(-B,A)$ we continue taking more data and update $L_t$; the first time $L_t$ exits $(-B,A)$ we stop (accepting more data) and we use the already accumulated information to decide between the two hypotheses $\Hyp_0,\Hyp_1$. If we call $\cS$ the time of stopping (which is clearly random, since it depends on the received data), then when $L_{\cS}\geq A$ we decide in favor of $\Hyp_1$, whereas if $L_{\cS}\leq-B$ in favor of $\Hyp_0$. The two thresholds $A,B$ are selected through simulations so that SPRT satisfies the two constraints in \eqref{eq:constr} with equality. This is always possible provided that $\alpha+\beta<1$. In the opposite case there is a trivial randomized test that can meet the two constraints without taking any samples (delay equal to 0). Note that these simulations to find proper values for $A,B$ are performed once offline, i.e. before the scheme starts, for each sensing environment.

The popularity of SPRT is due to its simplicity, but primarily to its very unique optimality properties. Regarding the latter we must say that optimality in the sense of \eqref{eq:optim}, \eqref{eq:constr} is assured only in the case of i.i.d.~observations. For more complex data models, SPRT is known to be only asymptotically optimum.

SPRT, when employed in our problem of interest, exhibits two serious practical weaknesses. First
the SUs need to send their local LLRs to the fusion center at the
Nyquist-rate of the signal; and secondly, the local LLR is a real
number which needs infinite (practically large) number of bits to be represented. These two
problems imply that a substantial communication overhead between the
SUs and the fusion center is incurred.
In this work, we are interested in \textit{decentralized} schemes
by which we mean that the SUs transmit
{\em low-rate} information to the fusion center.

\subsection{Decentralized Q-SPRT Scheme}
A straightforward way to achieve low-rate transmission is to let each SU transmit its local cumulative
LLR at a lower rate, say at time instants $T, 2T, ..., mT, ...$,
where the period $T \in {\mathbb N}$; and to quantize the local cumulative LLRs
using a finite number of quantization levels. Specifically, during time instants
$(m-1)T+1, ..., mT$, each SU computes its incremental
LLR $L^k_{mT}-L^k_{(m-1)T}$ of the observations $y^k_{(m-1)T+1}, \ldots, y^k_{mT}$,
to obtain
\begin{eqnarray} \label{unisam1.eq}
\lambda^k_{mT}\triangleq L^k_{mT}-L^k_{(m-1)T}=\sum_{t=(m-1)T+1}^{mT}l_t^k,
\end{eqnarray}
where $l_t^k$ is the LLR of observation $y_t^k$, defined in \eqref{eq:sampleLLR}.
It then quantizes $\lambda^k_{mT}$ into $\tilde{\lambda}^k_{mT}$ using a finite number $\tilde{r}$ of quantization levels. Although there are several ways to perform quantization, here we are going to analyze the simplest strategy, namely uniform quantization. We will also make the following assumption
\begin{equation}
\max_{k,t}|l_t^k|<\phi<\infty,
\label{eq:phi}
\end{equation}
stating that the LLRs of all observations are uniformly bounded by a finite constant $\phi$ across time and across SUs.

From \eqref{unisam1.eq} and \eqref{eq:phi} we can immediately conclude that $|\lambda^k_{mT}|< T\phi$. For our quantization scheme we can now divide the interval $(-T\phi,T\phi)$ uniformly into $\tilde{r}$ subintervals and assign the mid-value of each subinterval as the corresponding quantized value. Specifically we define
\begin{equation}
\tilde{\lambda}^k_{mT}=-T\phi+\frac{T\phi}{\tilde{r}}+\left\lfloor\frac{\tilde{r}(\lambda^k_{mT}+T\phi)}{2T\phi}\right\rfloor\frac{2T\phi}{\tilde{r}}.
\end{equation}
These quantized values are then transmitted to the FC. Of course the SU does not need to send the actual value but only its index which can be encoded with $\log_2\tilde{r}$ bits.

The FC receives the quantized information from all SUs, \textit{synchronously}, and updates the approximation of the global running LLR based on the information received, i.e.
\begin{eqnarray}\label{unisam2.eq}
{\tilde L}_{mT} = {\tilde L}_{(m-1)T} + \sum_{k=1}^K {\tilde \lambda}^k_{mT}.
\end{eqnarray}
Mimicking the SPRT introduced above, we can then define the following sequential scheme $(\tilde{\cS},\delta_{\tilde{\cS}})$, where
    \begin{align}
    \tilde{\cS}&=T\cM;~~\cM=\inf\left\{m>0: \tilde{L}_{mT}\not\in(-\tilde{B},\tilde{A})\right\};
  ~~\tilde{\delta}_{\tilde{\cS}} = \left\{ \begin{array}{ll}
	        1, & \mbox{if $\tilde{L}_{\tilde{\cS}}\geq \tilde{A}$}, \\
        0, & \mbox{if $\tilde{L}_{\tilde{\cS}}\leq -\tilde{B}$}.
        \end{array} \right.\label{eq:Q-SPRT}
    \end{align}
Again, the two thresholds $\tilde{A},\tilde{B}$ are selected to satisfy the two error probability constraints with equality. We call this scheme the \textit{Quantized-SPRT} and denote it as Q-SPRT.

As we will see in our analysis, the performance of Q-SPRT is directly related to the quantization error of each SU. Since we considered the simple uniform quantization, it is clear that
\begin{equation}
|\lambda^k_{mT}-\tilde{\lambda}^k_{mT}|<\frac{T\phi}{\tilde{r}}.
\label{eq:errQSPRT}
\end{equation}

We next consider three popular spectrum sensing methods and give the corresponding
local LLR expressions.

\subsection{Examples - Spectrum Sensing Methods}\label{sec:examples}

\subsubsection*{Energy Detector}
The energy detector performs spectrum sensing by detecting the primary user's
signal energy. We assume that when the primary user is present, the received
signal at the $k$-th SU is $y^k_{t} = x^k_t  + w^k_t$,
where $x^k_t$ is the received primary user signal, and
$w^k_t \sim {\cal N}_c(0,\sigma^2_w)$ is the
additive white Gaussian noise. Denote $\theta_k \triangleq \frac { \Exp[ |x^k_t|^2 ] } {\sigma^2_w/2}$
then the received signal-to-noise ratio (SNR) at the $k$-th SU is $\frac { \Exp[ |x^k_t|^2 ] } {\sigma^2_w}=\frac{\theta_k}{2}$.
Also define $\gamma^k_t \triangleq \frac { |y^k_t|^2 } {\sigma^2_w/2}$.
The energy detector is based on the following hypothesis testing formulation \cite{Ma_09}
\begin{eqnarray}
\begin{array} {cl}
\Hyp_0: & \gamma^k_t\sim \chi^2_2, \\
\Hyp_1: & \gamma^k_t\sim \chi^2_2(\theta_k),
\end{array}
\end{eqnarray}
where $\chi^2_2$ denotes a central chi-squared distribution with
2 degrees of freedom; and $\chi^2_2(\theta_k)$ denotes a non-central chi-squared distribution with
2 degrees of freedom and noncentrality parameter $\theta_k$.

Using the pdfs of central and non-central chi-squared distributions, we write the local
LLR, $l^k_t$, of the observations as follows
\be
l^k_t=\log \frac { \frac{1}{2} \exp\left(-\frac{\gamma^k_t+\theta_k}{2}\right) I_0\left(\sqrt{\theta_k \gamma^k_t}\right)}
                     { \frac{1}{2} \exp\left(-\frac{\gamma^k_t}{2}\right)}
=\log I_0\left(\sqrt{\theta_k \gamma^k_t}\right)-\frac{\theta_k}{2},
\ee
where $I_0(x)$ is the modified Bessel function of the first kind and $0$-th order.

\subsubsection*{Spectral Shape  Detector}
A certain class of primary user signals, such as the television broadcasting signals, exhibit
strong spectral correlation that can be exploited by the spectrum sensing algorithm \cite{Quan11}.
The corresponding hypothesis testing then consists in discriminating between the  channel's white
Gaussian noise, and the correlated primary user signal. The spectral shape of the primary user signal
is assumed known a priori, which can be approximated by a $p$-th order auto-regressive (AR) model.
Hence the hypothesis testing problem can be written as
\begin{equation}
\begin{array} {cl}
\Hyp_0: & y^k_{t}=w^k_t,  \\
\Hyp_1: & y^k_{t}=\sum_{i=1}^p a_i y^k_{t-i} + v^k_t,
\end{array}
\end{equation}
where $\{w^k_t\},\{v^k_t\}$ are i.i.d.~sequences with $w^k_t\sim {\cal N}_c (0, \sigma^2_w)$ and $v^k_t\sim {\cal N}_c (0, \sigma^2_v)$, while $a_1, \ldots, a_p$ are the AR model coefficients.

Using the Gaussian pdf the likelihoods under $\Hyp_0$ and $\Hyp_1$ can be easily derived.
Then, accordingly the local LLR of the sample received at time $t$ at the $k$-th SU can be written as
\begin{multline}
l^k_t= \log\frac{f^k_1(y^k_t|y^k_{t-1},\ldots,y^k_{t-p})}{f^k_0(y^k_t)}
= \log \frac { \frac 1 { \pi \sigma^2_v}  \exp\left[ - \frac 1 {\sigma^2_v}| y^k_t - \sum_{i=1}^p a_i y^k_{t-i} |^2 \right]}
{\frac 1 {\pi \sigma^2_w} \exp \left( - \frac 1 {\sigma^2_w} | y^k_t |^2 \right) } \\
= \frac 1 {\sigma^2_w} | y^k_t |^2 -
\frac 1 {\sigma^2_v} \Big| y^k_t - \sum_{i=1}^p a_i y^k_{t-i} \Big|^2 + \log \frac {\sigma^2_w} {\sigma^2_v}.
\end{multline}

\subsubsection*{Gaussian Detector}
In general, when the primary user is present, the received signal
by the $k$-th SU can be written as
$y^k_{t} = h^k_t s_t + w^k_t$, where $h^k_t \sim {\cal N}_c(0, \rho_k^2)$ is
the fading channel response between the primary user and
the $k$-th secondary user; $s_t$ is the
digitally modulated signal of the
primary user drawn from a certain modulation,
with $\Exp [|s_t|^2 ]= 1$;
and $w^k_t \sim {\cal N}_c(0,\sigma^2_w)$ is the
additive white Gaussian noise. It is shown
in \cite{Josep10} that under both fast fading
and slow fading conditions, spectrum sensing
can be performed based on the following
hypothesis testing between two   Gaussian signals:
\begin{equation}
\begin{array} {cl}
\Hyp_0: & y^k_{t} \sim {\cal N}_c (0, \sigma^2_w), \\
\Hyp_1: & y^k_{t} \sim {\cal N}_c (0,  \rho_k^2 + \sigma^2_w).
\end{array}
\end{equation}
Then, using the Gaussian pdf the local incremental LLR $l^k_t$ is derived as
\be
\label{eq:LLR1}
l^k_t=\log \frac {f^k_1 (y^k_{t} )} {f^k_0  (y^k_{t} )}= \log \frac { \frac 1 {\pi (\rho_k^2 + \sigma^2_w)} \exp \Big( - \frac { |y^k_{t}|^2 }{\rho_k^2+\sigma^2_w} \Big)}
{ \frac 1 {\pi   \sigma^2_w } \exp \Big( - \frac { |y^k_{t}|^2 }{ \sigma^2_w} \Big)}
=\frac{\rho_k^2}{\sigma^2_w(\rho_k^2+\sigma^2_w)}|y^k_{t}|^2 + \log \frac {\sigma^2_w}{\rho_k^2+\sigma^2_w}.
\ee

\section{Decentralized Spectrum Sensing via Level-triggered Sampling}\label{sec:ExplainScheme}
In this article, we achieve the low-rate transmission required by the decentralized SPRT by adopting
event-triggered sampling, that is, a sampling strategy in which the sampling times are dictated by the actual signal to be sampled, in a dynamic way and as the signal evolves in time. One could suggest to find the optimum possible combination of event-triggered sampling and sequential detection scheme by directly solving the double optimization problem defined in \eqref{eq:optim}, \eqref{eq:constr} over the triplet sampling, stopping time, and decision function. Unfortunately the resulting optimization turns out to be extremely difficult not accepting a simple solution. We therefore adopt an indirect path. In particular, we propose a decentralized spectrum sensing approach based on a simple form of event-triggered sampling, namely, the uniform level-triggered sampling. Then we show that the performance loss incurred by adopting this scheme as compared to the \textit{centralized} optimum SPRT is insignificant. This clearly suggests that solving the more challenging optimization problem we mentioned before, produces only minor performance gains.

\subsection{Uniform Level-triggered Sampling at each SU}
Using uniform level-triggered sampling, each SU
samples its local cumulative LLR process $\{L^k_t\}$ at a sequence of \textit{random times} $\{t_n^k\}$, which is particular to each SU. In other words we do not assume any type of synchronization in sampling and therefore communication. The corresponding sequence of samples is $\{L^k_{t^k_n}\}$ with the sequence of sampling times recursively defined as follows
\be
\label{eq:LSsamptime}
t_n^k \triangleq \inf\left\{t> t^k_{n-1}: L^k_t-L^k_{t^k_{n-1}} \not\in(-\Delta,\Delta)\right\},~t^k_0=0,~L^k_0=0.
\ee
where $\Delta$ is a constant. As we realize from \eqref{eq:LSsamptime} the sampling times depend on the actual \textit{realization} of the observed LLR process and are therefore, as we pointed out, random. Parameter $\Delta$ can be selected to control the \textit{average} sampling periods $\Exp_i[t^k_n-t^k_{n-1}],~i=0,1$.
In principle we would like the two average periods to coincide to some prescribed value $T$. For simplicity we will assume that the LLR of each observation is symmetric around its mean under the two hypotheses. This guarantees that the two average periods under the two hypotheses are the same. However, we are not going to assume that the observations have the same densities across SUs. This of course will make it impossible to assure that all SUs will communicate with the FC with the same average period if we use the same $\Delta$ at each SU, a property which is practically very desirable. In Section\,\ref{ssec:comp}, we propose a practically meaningful method to set this design parameter $\Delta$ in a way that assures a fair comparison of our method with the classical decentralized scheme, that is, Q-SPRT.

What is interesting with this sampling mechanism is that it is not necessary to know the
exact sampled value but only whether the incremental LLR $L^k_t-L^k_{t^k_{n-1}}$ crossed the upper or the lower threshold. This information can be represented by using a \textit{single} bit. Denote with $\{ b_n^k\}$ the sequence of these bits, where $b_n^k=+1$  means that the LLR increment crossed the upper boundary while $b_n^k=-1$ the lower. In fact we can also define this bit as $b_n^k=\text{sign}(\lambda^k_n)$ where $\lambda^k_n \triangleq L^k_{t_n^k}-L^k_{t^k_{n-1}}$.

We can now approximate the local incremental LLR as $\hat{\lambda}^k_n=b_n^k\Delta$, and since $L^k_{t^k_n}=\sum_{j=1}^n \lambda^k_j$  we conclude that we can approximate the local LLR at the sampling times using the following equation
\begin{eqnarray} \label{recover1.eq}
{\hat L}^k_{t_n^k} = \sum_{j=1}^n \hat{\lambda}^k_j = \sum_{j=1}^n   b^k_j \Delta={\hat L}^k_{t_{n-1}^k}+b_n^k\Delta.
\end{eqnarray}
Note that we have \textit{exact recovery}, i.e., ${\hat L}^k_{t_n^k}= L^k_{t_n^k}$, if the difference $L^k_t-L^k_{t^k_{n-1}}$, at the times of sampling, hits exactly one of the two boundaries $\pm\Delta$. This is for example the case when $\{L^k_t\}$ is a continuous-time process with continuous paths.

\ignore{
Since the amplitudes carry only 1-bit information it is clear that
the sampled information is encoded into the sampling instants and more precisely in the
inter-sampling intervals $\{\tau^k_n\}$ where
\begin{eqnarray}
 \tau_n^k &\triangleq& t_n^k - t_{n-1}^k.
\end{eqnarray}
In fact the pairs
$\{ (\tau_n^k, b_n^k) \}$ constitute the
complete information of the sampled signal since in addition
to (\ref{recover1.eq}) we can also recover the sampling times from
\begin{eqnarray}
 t_n^k = \sum_{j=1}^n \tau_j^k.
 \end{eqnarray}

Comparing this form of sampling with the conventional uniform-in-time sampling
it is seen that in the latter, the sampling instants (which are
integer multiples of the period $T$) bear no information and all
information of the sampled signal is encoded into the sampled
values $L^k_{mT}$. This is a very crucial difference with the level-triggered sampling
scheme which, as we pointed out, encodes all information in the
inter-sampling intervals instead of the sampled values.

 If we compare the two techniques in terms of number of bits
needed to represent the sampled signal, in general there is no visible difference.
Indeed, both representations are very similar since in the
classical uniform sampling the samples $L^k_{mT}$ are signed quantities
(amplitude plus sign bit) whereas in the level-triggered case in
the pair $\{ (\tau_n^k, b_n^k) \}$, the inter-sampling interval
$\tau_n^k$  is a positive quantity
and $b_n^k$ is a sign bit.
} 

The advantage of the level-triggered approach manifests itself if we desire to communicate the sampled information, as is the case of decentralized spectrum sensing. Indeed note that with classical sampling we need to transmit, every
 $T$ units of time, the \textit{real numbers} $L^k_{mT}$ (or their digitized version with fixed number of bits). On the other hand,
in the level-triggered case, transmission is performed at the random time instants $\{t_n^k\}$ and at each
$t_n^k$ we simply transmit the
\textit{1-bit} value $b_n^k$.
This property of 1-bit communication induces significant savings in bandwidth and
transmission power, which is especially valuable for the cognitive radio applications, where low-rate and
low-power signaling among the secondary users is a key issue for maintaining normal operating conditions
for the primary users.

We observe that by using (\ref{eq:recursive}), we have
$L^k_t-L^k_{t_{n-1}^k} = \sum_{j=t_{n-1}^k+1}^t l^k_j$ where, we recall, $l_t^k$ is the (conditional) LLR of the observation $y_t^k$ at time $t$ at the $k$th SU defined in \eqref{eq:sampleLLR}. Hence (\ref{eq:LSsamptime}) can be rewritten as
\be
\label{eq:LSsamptime2}
t_n^k = \inf\Big\{t> t_{n-1}^k: \sum_{j=t_{n-1}^k+1}^t l^k_j \not\in(-\Delta,\Delta)\Big\}.
\ee
The level-triggered sampling procedure at each secondary user is summarized in
Algorithm \ref{alg:SUalg}.
Until the fusion center terminates it, the algorithm produces the bit stream
 $\{b_n^k\}$ based on the local cumulative LLR values $L^k_t$ at time instants
 $\{ t_n^k\}$, and sends these bits to the fusion center instantaneously as they
 are generated.

\begin{algorithm}[h!]\small
\caption{\small The uniform level-triggered sampling procedure at the $k$-th SU}
\label{alg:SUalg}
\baselineskip=0.5cm
\begin{algorithmic}[1]
\STATE Initialization: $t \gets 0, \ \; n \gets 0$
\STATE $\lambda \gets 0$
\WHILE {$\lambda \in (-\Delta,\Delta)$}
    \STATE $t \gets t+1$
    \STATE Compute $l^k_t$ [cf. Sec.~ \ref{sec:examples}]
    \STATE $\lambda \gets \lambda + l^k_t$
\ENDWHILE
\STATE $n \gets n+1$
\STATE $t_n^k=t$
\STATE Send $b_n^k = {\rm sign}(\lambda)$ to the fusion center at time instant  $t_n^k$
\STATE Stop if the fusion center instructs so; otherwise go to line 2.
\end{algorithmic}
\end{algorithm}

\noindent
{\it Remarks:\ }
\bi
\item Note that the level-triggered sampling naturally censors unreliable local information gathered at SUs, and allows only informative LLRs to be sent to the FC.
\item Note also that each SU essentially performs a {\em local} SPRT with thresholds $\pm\Delta$.
    The stopping times of the local SPRT are the inter-sampling intervals 
    and the corresponding decisions are the  bits
    $\{ b_n^k\}$ where
    $b_n^k=+1$ and $b_n^k=-1$ favor $\Hyp_1$ and $\Hyp_0$
    respectively.
\ei

\subsection{Proposed Decentralized Scheme}
The bit streams  $\{b_n^k\}_k$ from different SUs arrive at the
FC asynchronously. Using (\ref{eq:1}) and (\ref{recover1.eq}),
the global running LLR at any time $t$ is approximated by
\begin{eqnarray}\label{hatL.eq}
{\hat L}_t = \sum_{k=1}^K {\hat L}^k_t = \Delta\sum_{k=1}^K \sum_{n: t_n^k \leq t}
 b_n^k.
\end{eqnarray}
In other words the FC adds all the received bits transmitted by all SUs up to time $t$ and then normalizes the result with $\Delta$. Actually the update of $\hat{L}_t$ is even simpler. If $\{t_n\}$ denotes the sequence of communication instants of the FC with any SU, and $\{b_n\}$ the corresponding sequence of received bits then ${\hat L}_{t_n}={\hat L}_{t_{n-1}}+b_n\Delta$ while the global running LLR is kept constant between transmissions. In case the FC receives more than one bit simultaneously (possible in discrete time), it processes each bit separately, as we described, following any random or fixed ordering of the SUs.

Every time the global LLR process $\{\hat{L}_t\}$ is updated at the FC it will be used in an SPRT-like test to decide whether to stop or continue (receiving more information from the SUs) and in the case of stopping to choose between the two hypotheses. Specifically the corresponding sequential test $(\hat{\cS},\hat{\delta}_{\hat{\cS}})$ is defined, similarly to the centralized SPRT and Q-SPRT, as
    \begin{align}
    \hat{\cS}=t_{\cN};~~\cN=\inf\left\{n>0: \hat{L}_{t_n}\not\in(-\hat{B},\hat{A})\right\};~~ 
  \hat{\delta}_{\hat{\cS}} =\left\{ \begin{array}{ll}
        1, & \mbox{if $\hat{L}_{\hat{\cS}}\geq \hat{A}$}, \\
        0, & \mbox{if $\hat{L}_{\hat{\cS}}\leq -\hat{B}$}.
        \end{array} \right.
    \label{eq:RLT-SPRT}
    \end{align}
$\hat{\cS}$ counts in physical time units, whereas $\cN$ in number of messages transmitted from the SUs to the FC. Clearly \eqref{eq:RLT-SPRT} is the equivalent of \eqref{eq:Q-SPRT} in the case of Q-SPRT and expresses the reduction in communication rate as compared to the rate by which observations are acquired. In Q-SPRT the reduction is deterministic since the SUs communicate once every $T$ unit times, whereas here it is random and dictated by the local level triggered sampling mechanism at each SU.
The thresholds $\hat{A},\hat{B}$, as before, are selected so that $(\hat{\cS},\hat{\delta}_{\hat{\cS}})$ satisfies the two error probability constraints with equality. The operations performed at the FC are also summarized in Algorithm\,\ref{alg:FCalg}.

\begin{algorithm}[h!]\small
\caption{\small The SPRT-like procedure at the fusion center}
\label{alg:FCalg}
\baselineskip=0.5cm
\begin{algorithmic}[1]
\STATE Initialization:   $\hat{L}\gets 0, \ \; n \gets 0$
\WHILE {$\hat{L} \in (-\hat{B},\hat{A})$}
    \STATE  $n \gets n+1$
    \STATE  Listen to the SUs and wait to receive the next bit $b_n$ at time $t_n$ from some SU
    \STATE  $ {\hat L} \gets {\hat L} + b_n \Delta$
\ENDWHILE
\STATE Stop at time $\hat{\cS}=t_n$
\IF {$\hat{L} \geq \hat{A}$}
    \STATE $\hat{\delta}_{\hat{\cS}}=1$ -- the primary user is present
\ELSE      
    \STATE $\hat{\delta}_{\hat{\cS}}=0$ -- the primary user is not present
\ENDIF
\STATE Inform all SUs the spectrum sensing result
\end{algorithmic}
\end{algorithm}

\ignore{
\noindent
{\it Remark:\ }
Note that each SU essentially performs a {\em local} SPRT with thresholds $\pm\Delta$.
    The stopping times of the local SPRT are the inter-sampling intervals 
    and the corresponding decisions are the  bits
    $\{ b_n^k\}$ where
    $b_n^k=+1$ and $b_n^k=-1$ favor $\Hyp_1$ and $\Hyp_0$
    respectively.
}

\subsection{Enhancement}\label{sec:enhance}
A very important source of performance degradation in our proposed scheme is the difference between the exact value of $L^k_t$ and its approximation $\hat{L}^k_t$ (see \cite{Fellouris11}). In fact the more accurately we approximate $L^k_t$ the better the performance of the corresponding SPRT-like scheme is going to be.
In what follows we discuss an enhancement to the decentralized spectrum sensing method described above at the
SU and FC, respectively. Specifically, for the SU, we consider using more than one bit to quantize the
local incremental LLR values, while at the FC, we are going to use this extra information in a specific reconstruction method that will improve the approximation $\hat{L}^k_t$ and, consequently, the approximation of the global running LLR. We anticipate that this enhancement will induce a significant improvement in the overall performance of the proposed scheme by using only a small number of additional bits. Finally we should stress that there is no need for extra bits in the case of continuous-time and continuous-path signals since, as we mentioned, in this case $\hat{L}^k_t$ and $L^k_t$ coincide.

\subsubsection*{Overshoot Quantization at the SU}
Recall that for the continuous-time case, at each sampling instant, either the upper or the
lower boundary can be hit exactly by the local LLR, and therefore the information transmitted
to the fusion center was simply
a 1-bit sequence and this is sufficient to recover completely the sampled LLR
using (\ref{recover1.eq}).
In discrete-time case, at the time of sampling, the
LLR is no longer necessarily equal to the boundary since, due to the discontinuity of the discrete-time signal, we can
overshoot the upper boundary or undershoot the lower boundary.
The over(under)shoot phenomenon introduces a discrepancy in the whole
system resulting in an additional information loss (besides the loss in time resolution due to sampling).
Here we consider the simple idea of  allowing the transmission of more than one bits, which could
help approximate more accurately the local LLR and consequently reduce the performance loss due to the over(under)shoot phenomenon.

Bit $b^k_n$ informs whether the difference $\lambda^k_n \triangleq L^k_{t_n^k}-L^k_{t_{n-1}^k}$ overshoots the upper threshold $\Delta$ or undershoots the lower threshold $-\Delta$. Consequently the difference
$q_n^k \triangleq |\lambda^k_n|-\Delta\ge0$,
corresponds to the absolute value of the over(under)shoot. It is exactly this value we intend to further quantize at each SU. Note that $q_n^k$ cannot exceed in absolute value the last observed LLR increment, namely $|l^k_{t^k_n}|$. To simplify our analysis we will assume that $|l^k_t| < \phi<\infty$ for all $k,t$ as in \eqref{eq:phi}. In other words the LLR of each observation is uniformly bounded across time and SUs.

Since for the amplitude $q_n^k$ of the over(under)shoot we have $q_n^k\le|l^k_{t^k_n}|$ this means that $0\le q_n^k\ < \phi$. Let us now divide the interval $[0,\phi)$, uniformly, into the following $\hat{r}$ subintervals $[(m-1),m)\frac{\phi}{\hat{r}},~m=1,\ldots,\hat{r}$. Whenever $q_n^k$ falls into one such subintervals the corresponding SU must transmit a quantized value $\hat{q}_n^k$ to the FC. Instead of adopting some deterministic strategy and always transmit the same value for each subinterval, we propose the following simple \textit{randomized} quantization rule
\be
\hat{q}_n^k=\left\{
\begin{array}{cl}
\lfloor\frac{q_n^k\hat{r}}{\phi}\rfloor\frac{\phi}{\hat{r}},&~~\text{with probability}~
p=\frac{1-\exp(q_n^k-(\lfloor\frac{q_n^k\hat{r}}{\phi}\rfloor+1)\frac{\phi}{\hat{r}})}{1-\exp(-\frac{\phi}{\hat{r}})}\\
(\lfloor\frac{q_n^k\hat{r}}{\phi}\rfloor+1)\frac{\phi}{\hat{r}},&~~\text{with probability}~
1-p=\frac{\exp(q_n^k-\lfloor\frac{q_n^k\hat{r}}{\phi}\rfloor\frac{\phi}{\hat{r}})-1}{\exp(\frac{\phi}{\hat{r}})-1}.
\end{array}
\right.
\label{eq:quant}
\ee
Simply said, if $q_n^k\in[(m-1),m)\frac{\phi}{\hat{r}}$ then we quantize $q_n^k$ either with the lower \textit{or} the upper end of the subinterval by selecting randomly between the two values. The quantized value $\hat{q}_n^k$ that needs to be transmitted to the FC clearly depends on the outcome of a random game and is not necessarily the same every time $q_n^k$ falls into the same subinterval. Regarding the randomization probability $p$ the reason it has the specific value depicted in \eqref{eq:quant} will become apparent in Lemma\,\ref{lem:quant}.

If we have $\hat{r}$ subintervals then we transmit $\hat{r}+1$ different messages corresponding to the values $m\frac{\phi}{\hat{r}},~m=0,\ldots,\hat{r}$. Combining them with the sign bit $b_n^k$ that also needs to be communicated to the FC, yields a total of $2(\hat{r}+1)$ possible messages requiring $\log_2 2(1+\hat{r})=1+\log_2(1+\hat{r})$ bits for transmitting this information. It is clear that each SU needs to transmit only an index value since we assume that the FC knows the list of all $2(1+\hat{r})$ quantized values.

\subsubsection*{Modified Update at the FC}
Let us now turn to the FC and see how the latter is going to use this additional information. Note that the $k$th SU, every time it samples, transmits the pair $(b_n^k,\hat{q}_n^k)$ where, we recall, the sign bit $b_n^k$ informs whether we overshoot the upper threshold $\Delta$ or undershoot the lower threshold $-\Delta$ and $\hat{q}_n^k$ the quantized version of the absolute value of the over(under)shoot. Consequently since we have $\lambda^k_n=b_n^k(\Delta+q_n^k)$ it is only natural to approximate the difference as follows
\begin{equation}
\hat{\lambda}^k_n=b_n^k\left(\Delta+\hat{q}_n^k\right),
\label{eq:increment}
\end{equation}
which leads to the following update of the local running LLR
\begin{equation}
\hat{L}^k_{t^k_n}=\hat{L}^k_{t^k_{n-1}}+b_n^k\left(\Delta+\hat{q}_n^k\right).
\end{equation}
This should be compared with the simplified version \eqref{recover1.eq} where the term $\hat{q}_n^k$ is missing. It is exactly this additional term that increases the accuracy of our approximation and contributes to a significant performance improvement in our scheme.
Of course the update of the global running LLR is much simpler since the FC, if it receives at time $t_n$ information $(b_n,\hat{q}_n)$ from some SU, then it will update its approximation of the global running LLR as follows
\begin{equation}
\hat{L}_{t_n}=\hat{L}_{t_{n-1}}+b_n\left(\Delta+\hat{q}_n\right).
\label{eq:LT-update}
\end{equation}
The updated value will be held constant until the next arrival of information from some SU.

For the SU operations given in Algorithm~\ref{alg:SUalg}, only line 10 should be modified when multiple bits
are used at each sampling instant, as follows

\begin{center}\small
 \begin{tabular}{ l  l} \hline
{\small 10:} & Quantize $q_n^k$ as in \eqref{eq:quant} and send $(b_n^k,\hat{q}_n^k)$ to the fusion center
at time  $t_n^k$.\\
\hline
\end{tabular}
\end{center}
On the other hand, for the FC operations given in Algorithm~\ref{alg:FCalg},
lines 4 and 5 should be modified as follows
\begin{center}\small
 \begin{tabular}{ l  l} \hline
{\small 4:} & Listen to the SUs and wait to receive the next message $(b_n,\hat{q}_n)$ from some SU. \\
{\small 5:} &  $ {\hat L} \gets {\hat L} + b_n(\Delta+\hat{q}_n)$. \\
\hline
\end{tabular}
\end{center}

With the proposed modification at each SU and at the FC we have in fact altered the communication protocol between the SUs and the FC and also the way the FC approximates the global running LLR. The final sequential test $(\hat{\cS},\hat{\delta}_{\hat{\cS}})$ however, is exactly the same as in \eqref{eq:RLT-SPRT}. We are going to call our decentralized test \textit{Randomized Level Triggered} SPRT and denote it as RLT-SPRT\footnote{In \cite{Fellouris11} the corresponding decentralized D-SPRT test that uses level triggered sampling at the sensors (that play the role of the SUs) is based only on 1-bit communication.}.
As we will demonstrate theoretically and also through simulations, the employment of extra bits in the communication between SUs and FC will improve, considerably, the performance of our test, practically matching that of the optimum.

Let us now state a lemma that presents an important property of the proposed quantization scheme.

\begin{lem}\label{lem:quant}\vskip-0.3cm
Let $\hat{q}_n^k$ be the $(\hat{r}+1)$-level quantization scheme defined in \eqref{eq:quant} for the overshoot $q_n^k$, then
\begin{gather}
\Exp[e^{\pm b_n^k(\Delta+\hat{q}_n^k)}|b_n^k,q_n^k]\leq e^{\pm b_n^k(\Delta+q_n^k)}= e^{\pm(L^k_{t^k_n}-L^k_{t^k_{n-1}})},\label{eq:lem1.1}
\end{gather}
where $\Exp[\cdot]$ denotes expectation with respect to the randomization probabilities.
\end{lem}

\begin{IEEEproof}
For given $q_n^k$, $\hat{q}^k_n$ takes the two values defined in \eqref{eq:quant} with probability $p$ and $1-p$ respectively. Define $\hat{\epsilon}=\frac{\phi}{\hat{r}}$, that is, the common length of the subintervals. Suppose that $(m-1)\hat{\epsilon}\leq q_n^k<m\hat{\epsilon},~m=1,\ldots,\hat{r}$ then $\hat{q}_n^k$ takes the two end values with probabilities $p,(1-p)$ respectively, but let us consider $p$ unspecified for the moment. We would like to select $p$ so that
\begin{equation}
pe^{\pm b_n^k(\Delta+(m-1)\hat{\epsilon})}+(1-p)e^{\pm b_n^k(\Delta+m\hat{\epsilon})}\leq e^{\pm b_n^k(\Delta+q_n^k)}.
\end{equation}
Since $b_n^k$ is a sign bit this is equivalent to solving the inequality
\begin{equation}
pe^{\pm (m-1)\hat{\epsilon}}+(1-p)e^{\pm m\hat{\epsilon}}\leq e^{\pm q_n^k},
\end{equation}
from which we conclude that
\begin{equation}
p=\min\left\{
\frac{e^{-m\hat{\epsilon}}-e^{-q^k_n}}{e^{-m\hat{\epsilon}}-e^{-(m-1)\hat{\epsilon}}},
\frac{e^{m\hat{\epsilon}}-e^{q^k_n}}{e^{m\hat{\epsilon}}-e^{(m-1)\hat{\epsilon}}}.
\right\}
\end{equation}
It is straightforward to verify that the second ratio is the smallest of the two, consequently we define $p$ to have this value which is the one depicted in \eqref{eq:quant}.
\end{IEEEproof}

Note that the approximation in the incremental LLR $L^k_{t^k_n}-L^k_{t^k_{n-1}}$ induces an equivalent approximation for the incremental LR $\exp(L^k_{t^k_n}-L^k_{t^k_{n-1}})$. The randomization is selected so that the latter, in average (over the randomization), does not exceed the exact incremental LR. One could instead select $p$ so that the average of the approximation of the incremental LLR matches the exact LLR value. Even though this seems as the most sensible selection, unfortunately, it leads to severe analytical complications which are impossible to overcome. The proposed definition of $p$, as we will see in the next section, does not have such problems.

\section{Performance Analysis}\label{sec:PerAn}
In this section, we provide an asymptotic analysis on the stopping time of the decentralized spectrum sensing
method based on the level-triggered sampling scheme proposed in Section~\ref{sec:ExplainScheme},
and compare it with the centralized SPRT procedure given by \eqref{eq:SPRT}. A similar comparison is performed for the conventional decentralized approach that uses uniform sampling and quantization
[cf. \eqref{unisam1.eq},\eqref{unisam2.eq}]. For our comparisons we will be concerned with the notion of asymptotic optimality for which we distinguish different levels \cite{Fellouris11},\cite{Moustakides11}.
\begin{dfn}\vskip-0.3cm
Consider any sequential scheme $(\cT,\delta_{\cT})$ with stopping time $\cT$ and decision function $\delta_{\cT}$ satisfying the two error probability constraints $\Pro_0(\delta_{\cT}=1)\le\alpha$ and $\Pro_1(\delta_{\cT}=0)\le\beta$. If $\cS$ denotes the optimum SPRT that satisfies the two error probability constraints with equality then,
as the Type-I and Type-II error probabilities $\alpha,\beta\to 0$,
the sequential scheme $(\cT,\delta_{\cT})$ is said to be order-1 asymptotically optimal if~\footnote{A quick reminder for the definitions of the notations $o(\cdot)$, $O(\cdot)$ and $\Theta(\cdot)$:
$f(x) = o\left(g(x)\right)$ if $f(x)$ grows with a lower rate than $g(x)$;
$f(x) = O\left(g(x)\right)$ if $f(x)$ grows with a rate that is no larger than the rate of $g(x)$; and
$f(x) = \Theta\left(g(x)\right)$ if $f(x)$ grows with exactly the same rate as $g(x)$.
Thus $o(1)$ represents a term that tends to 0. Particularly for this case we will write $o_x(1)$ to indicate a quantity that becomes negligible with $x$ and $o_{x,y}(1)$ to indicate a quantity that becomes negligible either with $x$ or with $y$ or with both.}
\be
\label{eq:asyopt1}
1\leq\frac{\Exp_i[\cT]}{\Exp_i[\cS]}=1+o_{\alpha,\beta}(1);
\ee
order-2 asymptotically optimal  if
\be
\label{eq:asyopt2}
0\leq\Exp_i[\cT]-\Exp_i[\cS]= O(1);
\ee
and finally order-3, if
\be
\label{eq:asyopt3}
0\leq\Exp_i[\cT]-\Exp_i[\cS]= o_{\alpha,\beta}(1),
\ee
where $\Pro_i(\cdot)$ and $\Exp_i[\cdot]$ denote probability and the corresponding expectation under hypothesis $\Hyp_i,~i=0,1$.
\end{dfn}

\noindent {\it Remark:} In our definitions the left-hand side inequalities are automatically satisfied because $\cS$ is the optimum test. Note that order-2 asymptotic optimality
implies order-1 because $\Exp_i[\cS],\Exp_i[\cT]\to\infty$ as $\alpha,\beta\to0$; the opposite is not necessarily true. Order-1 is the most frequent form of asymptotic optimality encountered in the literature but it is also the weakest. This is because it is possible $\Exp_i[\cT]$ to diverge from the optimum $\Exp_i[\cS]$ without bound and still have a ratio that tends to 1. Order-2 optimality clearly limits the difference to bounded values, it is therefore stronger than order-1. Finally the best would be the difference to become arbitrarily small, as the two error probabilities tend to 0, which is the order-3 asymptotic optimality. This latter form of asymptotic optimality is extremely rare in the Sequential Analysis literature and corresponds to schemes which, for all practical purposes, are considered as optimum per se.

Next we study the three sequential tests of interest, namely the optimum centralized SPRT, the Q-SPRT and the RLT-SPRT and compare the last two with the optimum in order to draw conclusions about their asymptotic optimality. We start by recalling from the literature the basic results concerning the tests of interest in continuous time. Then we continue with a detailed
presentation of the discrete-time case where we analyze the performance of Q-SPRT and RLT-SPRT when the corresponding quantization schemes have a number of quantization levels that depends on the error probabilities.

\subsection{Analysis of Centralized SPRT, Q-SPRT and RLT-SPRT}
With continuous-time and continuous-path observations at the SUs, it is known that RLT-SPRT, using only $1$-bit achieves \emph{order-2} asymptotic optimality \cite{Fellouris11}, whereas Q-SPRT cannot enjoy any type of optimality by using fixed number of bits \cite{Tsitsiklis93}.

In discrete time the corresponding analysis of the three sequential schemes of interest becomes more involved, basically due to the over(under)shoot effect. This is particularly apparent in RLT-SPRT where because of the over(under)shoots, 1-bit communication is no longer capable of assuring order-2 asymptotic optimality as in the continuous-time and continuous-path case. In order to recover this important characteristic in discrete time, we are going to use the enhanced quantization/communication scheme proposed in Section\,\ref{sec:enhance}. Let us now consider in detail each test of interest separately.

In discrete time, for the optimum centralized SPRT, we have the following lemma that provides the necessary information for the performance of the test.

\begin{lem}\label{lem:SPRT_d}
Assuming that the two error probabilities $\alpha,\beta\to0$ at the same rate, the centralized SPRT, $\cS$, satisfies
\begin{equation}
\Exp_0[\cS]\geq\frac{1}{K\I_0}\cH(\alpha,\beta)=\frac{|\log\beta|}{K\I_0}+o_{\beta}(1);~~
\Exp_1[\cS]\geq\frac{1}{K\I_1}\cH(\beta,\alpha)=\frac{|\log\alpha|}{K\I_1}+o_{\alpha}(1),
\label{eq:SPRTperf_d}
\end{equation}
where $\cH(x,y)=x\log\frac{x}{1-y}+(1-x)\log\frac{1-x}{y}$; and $\I_i=\frac{1}{K}|\Exp_i[L_1]|,~i=0,1$ are the average Kullback-Leibler information numbers of the process $\{L_t\}$ under the two hypotheses.
\end{lem}

\begin{IEEEproof}
It should be noted that these inequalities become equalities in the continuous-time continuous-path case. The proof can be found in \cite[Page 21]{Siegmund_85}.
\end{IEEEproof}

Let us now turn our attention to the two decentralized schemes, namely the classical Q-SPRT and the proposed RLT-SPRT.
We have the following theorem that captures the performance of Q-SPRT.

\begin{thm}\label{th:1}\vskip-0.3cm
Assuming that the two error probabilities $\alpha,\beta\to0$ at the same rate, and that the number $\tilde{r}$ of quantization levels increases with $\alpha,\beta$, then the performance of Q-SPRT, $\tilde{\cS}$, as compared to the optimum centralized SPRT, $\cS$, satisfies
\begin{align}
\begin{split}
0&\le\Exp_1[\tilde{\cS}]-\Exp_1[\cS]\le \frac{2|\log\alpha|}{K\I_1^2}\frac{\phi}{\tilde{r}}\{1+o_{\tilde{r}}(1)\}
+T\frac{\phi}{\I_1}\{1+o_{\tilde{r}}(1)\}+o_{\alpha}(1)\\
0&\le\Exp_0[\tilde{\cS}]-\Exp_0[\cS]\le \frac{2|\log\beta|}{K\I_0^2}\frac{\phi}{\tilde{r}}\{1+o_{\tilde{r}}(1)\}
+T\frac{\phi}{\I_0}\{1+o_{\tilde{r}}(1)\}+o_{\beta}(1).
\end{split}
\label{eq:th1}
\end{align}
\end{thm}
\begin{IEEEproof}
The proof can be found in Appendix A.
\end{IEEEproof}

As with the classical scheme, let us now examine the behavior of the proposed test when the number of quantization levels increases as a function of the two error probabilities $\alpha,\beta$. We have the next theorem that summarizes the behavior of RLT-SPRT.

\begin{thm}\label{th:2}\vskip-0.3cm
Assuming that the two error probabilities $\alpha,\beta\to0$ at the same rate, and that the number $\hat{r}$ of quantization levels increases with $\alpha,\beta$, then the performance of RLT-SPRT, $\hat{\cS}$, as compared to the optimum centralized SPRT, $\cS$, satisfies
\begin{align}
\begin{split}
0&\le\Exp_1[\hat{\cS}]-\Exp_1[\cS]\leq \frac{|\log\alpha|}{K\I_1\Delta\tanh(\frac{\Delta}{2})}\frac{\phi}{\max\{\hat{r},1\}}\{1+o_{\Delta,\hat{r}}(1)\}+\frac{1}{\I_1}(\Delta+\phi)+o_{\Delta,\hat{r}}(1)+o_{\alpha}(1),\\
0&\le\Exp_0[\hat{\cS}]-\Exp_0[\cS]\leq \frac{|\log\beta|}{K\I_0\Delta\tanh(\frac{\Delta}{2})}\frac{\phi}{\max\{\hat{r},1\}}\{1+o_{\Delta,\hat{r}}(1)\}+\frac{1}{\I_0}(\Delta+\phi)+o_{\Delta,\hat{r}}(1)+o_{\beta}(1).
\end{split}
\label{eq:th2}
\end{align}
\end{thm}
\begin{IEEEproof}
The proof is presented in Appendix B.
\end{IEEEproof}

\subsection{Comparisons}\label{ssec:comp}
In order to make fair comparisons, the two decentralized schemes need to satisfy the same communication constraints. First, each SU is allowed to use at most $s$ bits per communication. This means that the number of quantization levels $\tilde{r}$ in Q-SPRT must satisfy $\tilde{r}=2^s$ while for RLT-SPRT we have $2(1+\hat{r})=2^s$ suggesting that $\hat{r}=2^{s-1}-1$.

The second parameter that needs to be specified is the information flow from the SUs to the FC. Since receiving more messages per unit time increases the capability of the FC to make a  faster decision, it makes sense to use the \textit{average rate of received messages} by the FC as a measure of the information flow.
In Q-SPRT, every $T$ units of time the FC receives, synchronously, $K$ messages (from all SUs), therefore the average message rate is $\frac{K}{T}$. Computing the corresponding quantity for RLT-SPRT is less straightforward. Consider the time interval $[0,t]$ and denote with $\cN_t$ the total number of messages received by the FC until $t$. We clearly have $\cN_t=\sum_{k=1}^K\cN_t^k$ where $\cN_t^k$ is the number of messages sent by the $k$th SU. We are interested in computing the following limit
\begin{equation}
\lim_{t\to\infty}\frac{\cN_t}{t}=\lim_{t\to\infty}\sum_{k=1}^K\frac{\cN^k_t}{t}=\sum_{k=1}^K\lim_{t\to\infty}\frac{1}{\frac{1}{\cN^k_t}(\sum_{n=1}^{\cN^k_t}t^k_n-t^k_{n-1})+\frac{1}{\cN^k_t}(t-t^k_{\cN^k_t})}=\sum_{k=1}^K\frac{1}{\Exp_i[t^k_1]},
\end{equation}
where we recall that $\{t^k_n\}$ is the sequence of sampling times at the $k$th SU and for the last equality we used the Law of Large Numbers since when $t\to\infty$ we also have $\cN^k_t\to\infty$. Consequently we need to select $\Delta$ so that $\sum_{k=1}^K\frac{1}{\Exp_i[t^k_1]}=\frac{K}{T}$. To obtain a convenient formula we are going to become slightly unfair for RLT-SPRT. From \eqref{eq:lem.period1} in Lemma\,\ref{lem:period} we have that $\frac{1}{\Exp_i[t_1^k]}\leq\frac{|\Exp_i[L_1^k]|}{\Delta\tanh(\frac{\Delta}{2})}$, which means that, $\sum_{k=1}^K\frac{1}{\Exp_i[t_1^k]}\leq\frac{K\I_i}{\Delta\tanh(\frac{\Delta}{2})}$. Therefore, if we set $\frac{K\I_i}{\Delta\tanh(\frac{\Delta}{2})}=\frac{K}{T}$ or, equivalently, $\Delta\tanh(\frac{\Delta}{2})=T\I_i$, the average message rate of RLT-SPRT becomes slightly smaller than the corresponding of Q-SPRT. Note that the average Kullback-Leibler information numbers, $\I_i, i=0,1$, can be once computed offline via simulations.

Under the previous parameter specifications, we have the following final form for the performance of the two schemes. For Q-SPRT
\begin{align}
\begin{split}
0&\le\Exp_1[\tilde{\cS}]-\Exp_1[\cS]\le \frac{\phi}{K\I_1^2}\frac{|\log\alpha|}{2^{s-1}}\{1+o_{s}(1)\}
+T\frac{\phi}{\I_1}\{1+o_{s}(1)\}+o_{\alpha}(1)\\
0&\le\Exp_0[\tilde{\cS}]-\Exp_0[\cS]\le \frac{\phi}{K\I_0^2}\frac{|\log\beta|}{2^{s-1}}\{1+o_{s}(1)\}
+T\frac{\phi}{\I_0}\{1+o_{s}(1)\}+o_{\beta}(1);
\end{split}
\label{eq:th11}
\end{align}
while for RLT-SPRT
\begin{align}
\begin{split}
0&\le\Exp_1[\hat{\cS}]-\Exp_1[\cS]\leq\frac{1}{T}\frac{\phi}{K\I_1^2}\frac{|\log\alpha|}{\max\{2^{s-1}-1,1\}}\{1+o_{T,s}(1)\}+T +\frac{\phi}{\I_1}+o_{T,s}(1)+o_{\alpha}(1),\\
0&\le\Exp_0[\hat{\cS}]-\Exp_0[\cS]\leq\frac{1}{T}\frac{\phi}{K\I_0^2}\frac{|\log\beta|}{\max\{2^{s-1}-1,1\}}\{1+o_{T,s}(1)\}+T +\frac{\phi}{\I_0}+o_{T,s}(1)+o_{\alpha}(1).
\end{split}
\label{eq:th22}
\end{align}

Comparing \eqref{eq:th11} with \eqref{eq:th22} there is a definite resemblance between the two cases. However in RLT-SPRT we observe the factor $\frac{1}{T}$ in the first term of the right hand side which, as we will immediately see, produces significant performance gains. Since $T$ is the communication period, and we are in discrete time, we have $T\geq1$. Actually, for the practical problem of interest we have $T\gg1$ suggesting that the first term in RLT-SPRT is smaller by a factor $T$, which can be large.

\ignore{
We selected the parameters of the two schemes so that the corresponding average message rate is equal to $\frac{K}{T}$ which means that the average \textit{bit} rate is $\frac{Ks}{T}$. Note that $\frac{Ks}{T}=K\frac{(s/D)}{(T/D)}=K\frac{sD}{TD}$ where $D$ any positive integer. This means that we can have the same average bit rate by reducing the number of bits by a factor $D$ and increasing the communication rate by the same factor $D$ or increasing the number of bits by a factor $D$ and reduce the communication rate by the same factor. Consequently, if we are allowed to have a maximal bit rate, the question is whether it is preferable to have 1-bit and frequent  or several bits and infrequent communications?
}

For fixed $T$ and $s$, none of the two schemes is asymptotically optimum even of order-1. However, in RLT-SPRT we can have order-1 asymptotic optimality when we fix the number of bits $s$ and impose large communication periods. Indeed using \eqref{eq:SPRTperf_d} of Lemma\,\ref{lem:SPRT_d} we obtain
\begin{equation}
\label{eq:order-1}
1\leq\frac{\Exp_1[\hat{\cS}]}{\Exp_1[\cS]}=1+\frac{\Exp_1[\hat{\cS}]-\Exp_1[\cS]}{\Exp_1[\cS]}\leq1+\Theta\left(\frac{1}{T}\right)+\frac{T K\I_1}{|\log\alpha|}+o_\alpha(1),
\end{equation}
consequently selecting $T\to\infty$ but $\frac{T}{|\log\alpha|}\to0$ we assure order-1 optimality. It is easy to verify that the best speed of convergence towards 1 of the previous right hand side expression is achieved when $T=\Theta(\sqrt{|\log\alpha|})$.

We should emphasize that similar order-1 optimality result, just by controlling the period $T$, cannot be obtained in Q-SPRT, and this is due to the missing factor $\frac{1}{T}$ in \eqref{eq:th11}. Consequently this is an additional indication (besides the continuous-time case) that the proposed scheme is more efficient than the classical decentralized Q-SPRT.

Let us now examine how the asymptotic optimality properties of the two methods improve when we allow the number of bits $s$ to grow with $\alpha,\beta$, while keeping $T$ constant. Note that in the case of Q-SPRT selecting $2^{s-1}=|\log\alpha|$ or, equivalently, $s=1+\log_2|\log\alpha|$ assures order-2 asymptotic optimality. For RLT-SPRT, using for simplicity the approximation $2^{s-1}-1\approx2^{s-1}$, the same computation yields $s=1+\log_2|\log\alpha|-\log_2T$. Of course the two expressions are of the same order of magnitude, however in RLT-SPRT the additional term $-\log_2T$, for all practical purposes, can be quite important resulting in a need of significantly less bits than Q-SPRT to assure order-2 asymptotic optimality. The conclusions obtained through our analysis, as we will see in the next section, are also corroborated by our simulations.

\section{Simulation Results}\label{sec:Sim}
In this section, we provide simulation results  to evaluate the
performance of the proposed cooperative spectrum sensing technique
based on level-triggered sampling and that
based on conventional uniform sampling, and how the two tests compare with the optimum centralized
scheme. In the simulations, the sampling period of the uniform
sampling is set as $T=4$. For the level triggered sampling, we adjust
the local threshold $\Delta$ so that the average rate of received messages by the FC
matches that of uniform sampling, i.e. $\sum_{k=1}^K \frac{1}{\Exp_i[t^k_1]}=\frac{K}{T}$
(see Section \ref{ssec:comp}).
The upper-bound $\phi$ for overshoot values is set as the $10^{-4}$ quantile of the LLR of a single observation which is computed once offline via simulations.
We mainly consider a cognitive radio system with two SUs, i.e., $K=2$, but
the effect of increasing user diversity is also analyzed in the last subsection.
All results are obtained by averaging
$10^4$ trials and using importance sampling to compute probabilities of rare events. We primarily focus on the energy detector since it is the most widely used
spectrum sensing method. The results
for the spectral shape detector and the Gaussian detector are quite similar.
In the subsequent figures average sensing delay performances are plotted under $\Hyp_1$.

\textit{\underline{Fixed SNR and $K$, varying $\alpha,\beta$}}: We first verify the theoretical findings presented in Section \ref{sec:PerAn}
\begin{figure}
\centering
\includegraphics[scale=0.85]{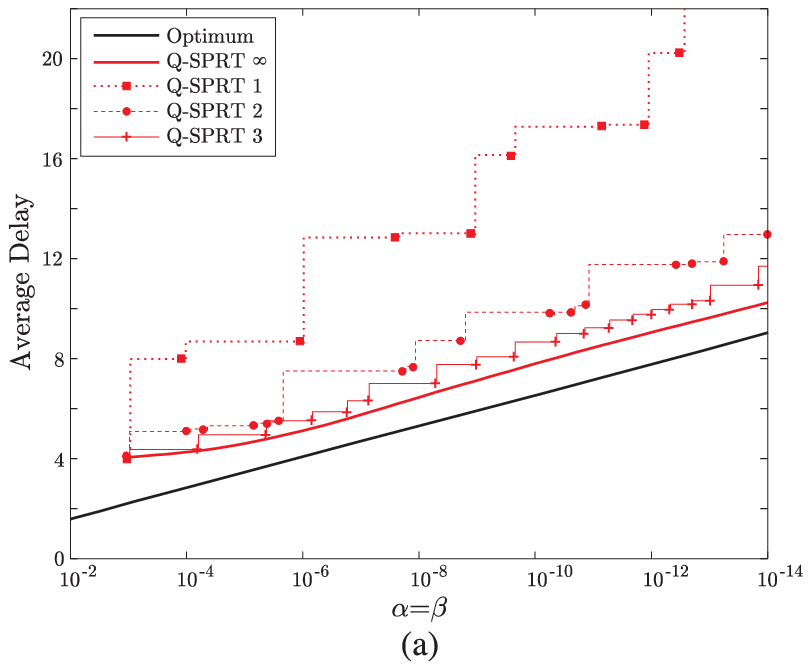}~~\includegraphics[scale=0.85]{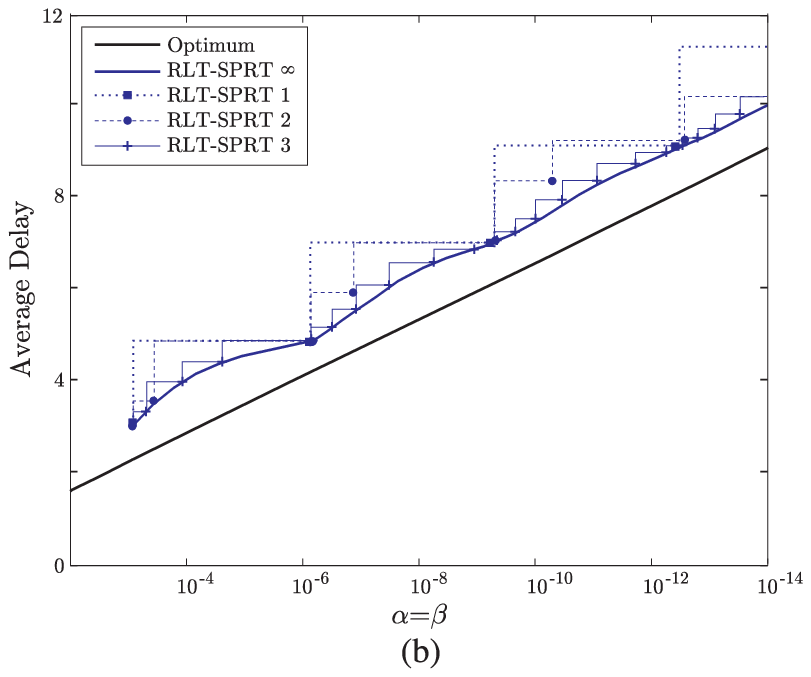}
\vskip0.2cm
\includegraphics[scale=0.85]{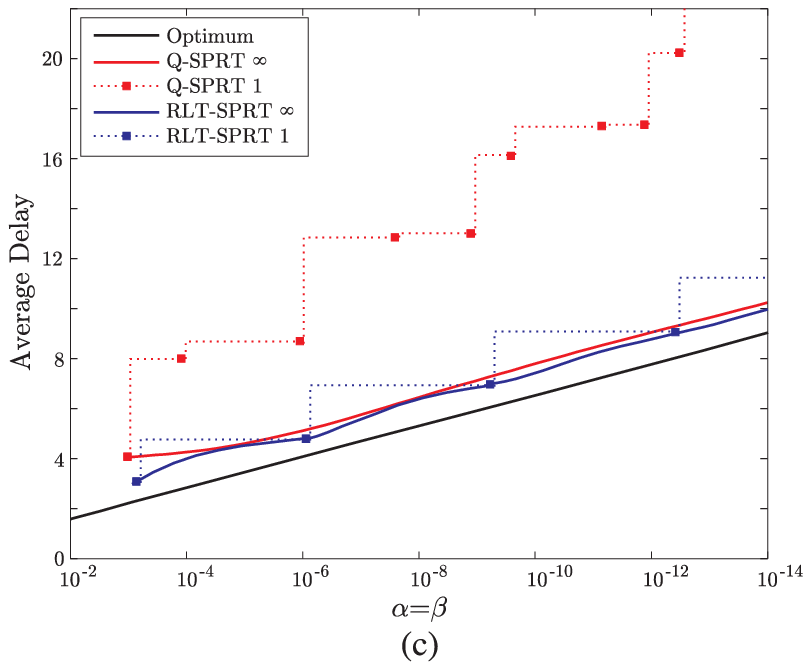}
\caption{Average detection delay vs error probabilities ($\alpha,\beta$) for optimum centralized and Q-SPRT, RLT-SPRT with 1,2,3,$\infty$ number of bits.}
\label{fig:asy}
\end{figure}
on the asymptotic optimality properties of the decentralized schemes.
We assume two SUs operate in the system, i.e. $K=2$.
For the energy detector, we set the receiver SNR for each SU to
$5$ dB and vary  the error probabilities $\alpha$ and $\beta$ together between
$10^{-1}$ and $10^{-10}$.

Fig.~\ref{fig:asy} illustrates asymptotic performances of the decentralized schemes using
1,2, 3 and $\infty$ number of bits.
Our first interesting result is the fact that by using a finite number of bits we can only achieve a discrete number of error rates.
Specifically, if a finite number of bits is used to represent local incremental LLR packages, then there is a finite number of possible values to update the global running LLR
(e.g. for one bit we have $\pm\Delta$).
Hence, the global running LLR, which is our global test statistic, can assume only a discrete number
of possible values. This suggests that any threshold between two consecutive LLR values will produce the same error probability.  Consequently, only a discrete set of
error probabilities ($\alpha,\beta$) are achievable.
Increasing the number of bits clearly increases the number of available error probabilities.
With infinite number of bits any error probability can be achieved.
The case of infinite number of bits corresponds to the best achievable performance for
Q-SPRT and RLT-SPRT.
Having their performance curves parallel to that of the optimum centralized scheme,
the $\infty$-bit case for both Q-SPRT and RLT-SPRT exhibits order-2 asymptotic optimality.
Recall that both schemes can enjoy order-2 optimality if the number of bits tends to infinity
with a rate of $\log|\log \alpha|$.

It is notable that the performance of RLT-SPRT with a small number of bits is very close to that
of $\infty$-bit RLT-SPRT at achievable error rates.
For instance, the performance of $1$-bit case coincides with that of $\infty$-bit case, but only
at a discrete set of points as can be seen in Fig.\,\ref{fig:asy}--b. However, we do not observe this feature for Q-SPRT.
Q-SPRT with a small number of bits (especially one bit) performs significantly worse than $\infty$-bit case Q-SPRT as well as its RLT-SPRT counterpart.
In order to achieve a target error probability that is not in the achievable set of a specific finite
bit case, one should use the thresholds corresponding to the closest smaller error
probability. This incurs a delay penalty in addition to the delay of the $\infty$-bit
case for the target error probability, demonstrating the advantage of using more bits.
Moreover, it is a striking result that $1$-bit RLT-SPRT is superior to $\infty$-bit Q-SPRT at
its achievable error rates, which can be seen in Fig.\,\ref{fig:asy}--c.

\begin{figure}[!ht]
\centering
\includegraphics[scale=0.85]{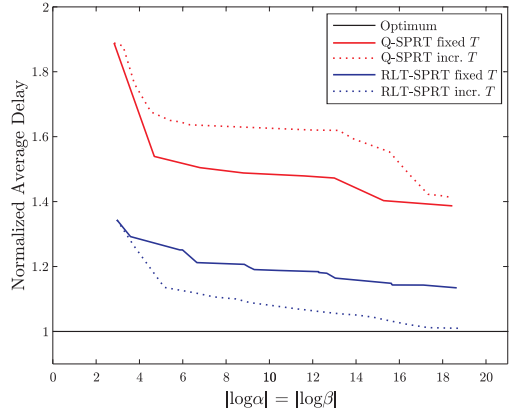}
\caption{Average detection delay normalized by the optimum centralized performance vs. error probabilities ($\alpha,\beta$) for Q-SPRT and RLT-SPRT with 2 bits and communication period either $T=4$ or $T=\Theta(\sqrt{|\log \alpha|})$.}
\label{fig:D}
\end{figure}

Fig.\,\ref{fig:D} corroborates the theoretical result related to order-1 asymptotic optimality that is obtained in (\ref{eq:order-1}). Using a fixed a number of bits, $s=2$, the performance of RLT-SPRT improves and achieves order-1 asymptotic optimality, i.e. $\frac{\Exp_1[\hat{\cS}]}{\Exp_1[\cS]}=1+o(1)$, as the communication period tends to infinity, $T=\Theta(\sqrt{|\log \alpha|})$. Conversely, the performance of Q-SPRT deteriorates under the same conditions.  Although in both cases Q-SPRT converges to the same performance level, its convergence speed is significantly smaller in the increasing $T$ case, which can be obtained theoretically by applying the derivation in (\ref{eq:order-1}) to (\ref{eq:th11}). This important advantage of RLT-SPRT over Q-SPRT is due to the fact that the
quantization error (overshoot error) observed by SUs at each communication time in RLT-SPRT depends only on the LLR of a single observation, but not on the communication period, whereas that in Q-SPRT increases with increasing communication period. Consequently, quantization error accumulated at the FC becomes smaller in RLT-SPRT, but larger in Q-SPRT when $T=\Theta(\sqrt{|\log \alpha|})$ compared to the fixed $T$ case. Note in Fig.\,\ref{fig:D} that, as noted before, only a discrete number of error rates are achievable since two bits are used. Here, we preferred to linearly combine the achievable points to emphasize the changes in the asymptotic performances of RLT-SPRT and Q-SPRT although the true performance curves of the $2$-bit case should be stepwise as expressed in Fig.\,\ref{fig:asy}.

\begin{figure}[!ht]
\centering
\includegraphics[scale=0.85]{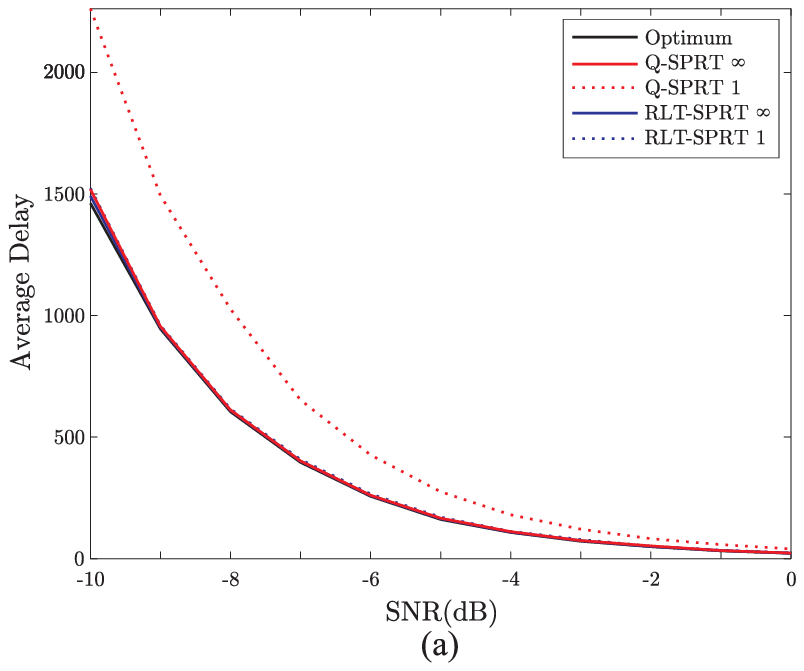}~~~~~\includegraphics[scale=0.85]{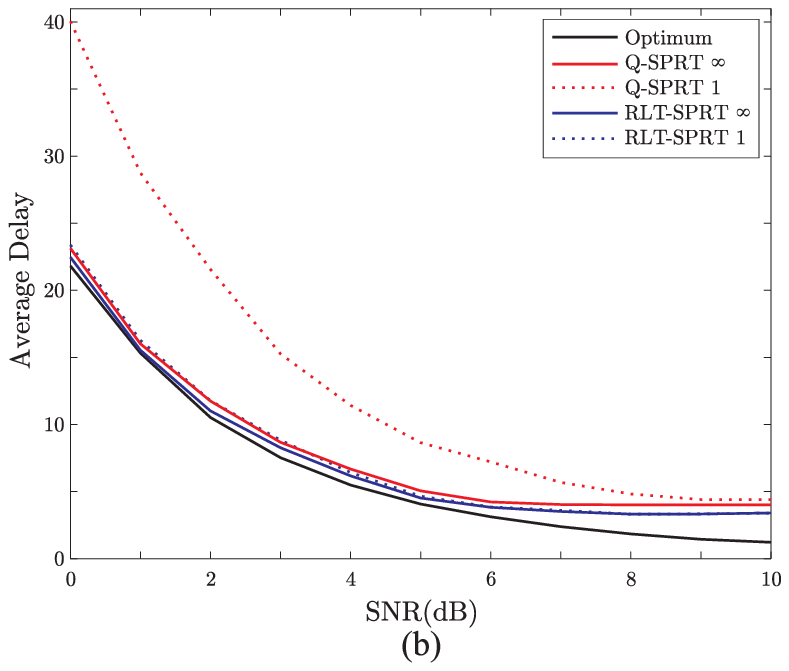}
\caption{Average detection delay vs SNR for optimum centralized and Q-SPRT, RLT-SPRT with 1,$\infty$ number of bits.}
\label{fig:snr}
\end{figure}

\textit{\underline{Fixed $\alpha, \beta$ and $K$, varying SNR}}:
Next, we consider the sensing delay performances of Q-SPRT and RLT-SPRT
under different SNR conditions with fixed $\alpha=\beta =10^{-6}$ and $K=2$.
In Fig.\,\ref{fig:snr}, it is clearly seen that at low SNR values there is a huge difference
between Q-SPRT and RLT-SPRT when we use one bit, which is the most important case in practice.
This remarkable difference stems from the fact that the one bit RLT-SPRT transmits the most
part of the sampled LLR information (except the overshoot), whereas Q-SPRT fails to transmit
sufficient information by quantizing the LLR information.
Moreover, as we can see the performance of the $1$-bit RLT-SPRT is very close to that of the infinite bit case and the optimum centralized scheme.
At high SNR values depicted in Fig.\,\ref{fig:snr}--b, schemes all behave similarly, but again RLT-SPRT is superior to Q-SPRT.
This is because the sensing delay of Q-SPRT cannot go below the sampling interval $T=4$,
whereas RLT-SPRT is not bounded by this limit due to the asynchronous communication it implements.

\begin{figure}[!ht]
\centering
\includegraphics[scale=0.85]{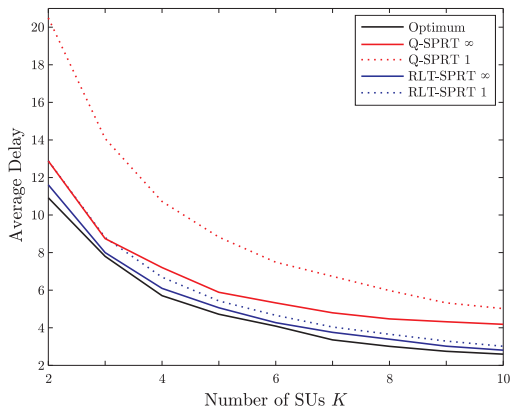}
\caption{Average detection delay vs number of SUs ($K$) for optimum centralized and Q-SPRT, RLT-SPRT with 1,$\infty$ number of bits.}
\label{fig:K}
\end{figure}

\textit{\underline{Fixed SNR, $\alpha$ and $\beta$, varying $K$}}:
We, then, analyze the case where the user diversity increases.
In Fig.\,\ref{fig:K}, it is seen that with increasing number of SUs, the average sensing delays of all schemes decay with the same rate of $1/K$ as shown in Section~\ref{sec:PerAn} (cf. (\ref{eq:SPRTperf_d}), (\ref{eq:th1}) and (\ref{eq:th2})).
The decay is more notable for the $1$-bit case because the overshoot accumulation is more intense, but very quickly becomes less pronounced as we increase the number of SUs.
It is again interesting to see that the $1$-bit RLT-SPRT is superior to the $\infty$-bit Q-SPRT for values of $K$ greater than $3$.

\begin{figure}[!ht]
\centering
\includegraphics[scale=0.4]{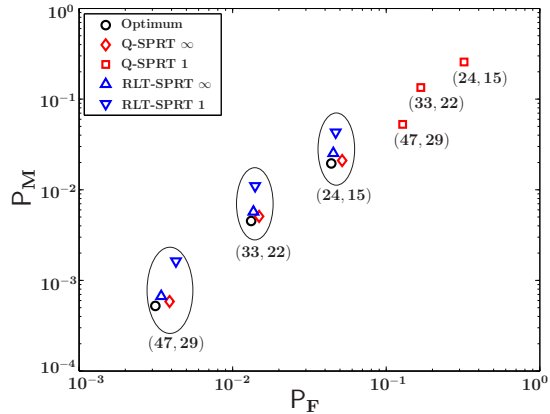}
\caption{False alarm probability ($\Pro_\text{F}$) vs. misdetection probability ($\Pro_\text{M}$) for optimum centralized and Q-SPRT, RLT-SPRT with 1,$\infty$ number of bits.}
\label{fig:roc}
\end{figure}

\textit{\underline{Fixed SNR, $K$ and $\Exp_i[\cT]$, varying $\Pro_\text{F}$ and $\Pro_\text{M}$}}:
Finally, we show the operating characteristics of the schemes for fixed SNR$=-3$ dB, $K=2$ and $\Exp_i[\cT]$.
False alarm ($\Pro_\text{F}$) and misdetection ($\Pro_\text{M}$) probabilities of the schemes are given in Fig.\,\ref{fig:roc} when they have exactly the same three average delay pairs, $(\Exp_0[\cT],\Exp_1[\cT])$.
Note that no target error probabilities,$\alpha$ and $\beta$, are specified in this set of simulations. Thresholds, $A$ and $B$, are dictated by the given average delays. 
We again observe here that $1$-bit Q-SPRT performs considerably worse than $1$-bit RLT-SPRT and the other schemes. Its error probability pairs are far away from the group of others. On the other hand, error probability pairs of $1$-bit RLT-SPRT are close to those of the $\infty$-bit schemes and the optimum centralized scheme.

\section{Conclusion}

We have proposed and rigorously analyzed a new spectrum sensing scheme for cognitive radio networks.
The proposed scheme is based on \emph{level-triggered sampling} which is a non-uniform sampling technique that naturally outputs $1$ bit information without performing any 	quantization, and allows SUs
to communicate to the FC asynchronously.
Therefore, it is truly decentralized, and it ideally suits the cooperative spectrum sensing in
cognitive radio networks.
With continuous-time observations at the SUs our scheme  achieves \emph{order-2} asymptotic optimality by using only $1$ bit. However, its conventional uniform sampling
counterpart Q-SPRT cannot achieve any type of optimality by using any fixed number of bits.
With discrete-time observations at the SUs our scheme achieves \emph{order-2} asymptotic optimality by means of an additional randomized quantization step (RLT-SPRT) when the number of bits tends to infinity at a considerable slow rate, $O(\log|\log \alpha|)$.
In particular, RLT-SPRT needs significantly less number of bits to achieve order-2 optimality than Q-SPRT.
With a fixed number of bits, unlike Q-SPRT our scheme can also attain \emph{order-1} asymptotic optimality when the average communication period tends to infinity at a slower rate than $|\log \alpha|$.

Simulation results showed that with a finite number of bits only a discrete set of error probabilities are available due to the updating of the global running LLR with a finite number of possible values.
RLT-SPRT, using $1$ bit, performs significantly better than $1$-bit Q-SPRT, and even
better than $\infty$-bit Q-SPRT at its achievable error rates.
We also provided simulation results for varying SNR conditions and increasing SU diversity.
Using $1$ bit RLT-SPRT performs remarkably better than $1$-bit Q-SPRT.
It also attains the performance of $\infty$-bit Q-SPRT at low SNR values, and even outperforms $\infty$-bit Q-SPRT for SNR greater than $3$ dB or when the number of SUs exceeds $3$.

\ignore{
Finally we must emphasize the important practical advantage of RLT-SPRT not requiring synchronization, unlike Q-SPRT, and its capability to censor uninformative SUs by reducing their communication rate with the fusion center.}

\abovedisplayskip=0.15cm
\belowdisplayskip=0.15cm
\section*{Appendix A}
In this appendix we are going to prove the validity of Theorem\,\ref{th:1}. Let us first introduce a technical lemma.

\begin{lem}\label{lem:appA.1}\vskip-0.3cm
If $\tilde{A},\tilde{B}$ are the thresholds of Q-SPRT with sampling period $T$ and $\tilde{r}$ quantization levels, then for sufficiently large $\tilde{r}$ we have
\begin{equation}
\tilde{A}\leq\frac{|\log\alpha|}{\rho_0};~~~
\tilde{B}\leq\frac{|\log\beta|}{\rho_1},
\label{eq:appA.1}
\end{equation}
where $\rho_0,\rho_1>0$ are the solutions of the equations $\Exp_0[e^{\rho_0\tilde{L}_T}]=1$ and $\Exp_1[e^{-\rho_1\tilde{L}_T}]=1$ respectively. Furthermore we have $\rho_i\geq1-\frac{\phi}{\I_i\tilde{r}}\{1+o_{\tilde{r}}(1)\}$ where, we recall, $\I_i=\frac{1}{K}|\Exp_i[L_1]|$ and $\phi$ is the maximum of the absolute LLR $|l_1^k|$ of a single observation at any SU, as defined in \eqref{eq:phi}.
\end{lem}
\begin{IEEEproof}
We will only show the first inequality in \eqref{eq:appA.1} since the second can be show in exactly the same way.
We recall that the two thresholds $\tilde{A},\tilde{B}$ are selected so that the two error probabilities are satisfied with equality. In particular we have $\alpha=\Pro_0(\tilde{L}_{\tilde{\cS}}\ge\tilde{A})$.

From the definition of the stopping time in \eqref{eq:SPRT} we have $\tilde{\cS}=T\cM$, that is, $\tilde{\cS}$ is an integer multiple of the period $T$. Note now that $\tilde{L}_{mT}=\sum_{n=1}^m\tilde{\lambda}_{nT}$ where $\tilde{\lambda}_{nT}=\sum_{k=1}^K\tilde{\lambda}^k_{nT}$. Since we have independence across time and SUs we conclude that the sequence $\{\tilde{\lambda}_{nT}\}$ is i.i.d.~under both hypotheses.

Let $\rho_0$ be the solution to the equation $\psi(\rho)=\Exp_0[e^{\rho\tilde{L}_T}]=\Exp_0[e^{\rho\tilde{\lambda}_T}]=1$, with the second equality being true because $\tilde{L}_T=\tilde{\lambda}_T$. It is easy to see that $\psi(\rho)$ is a convex function of $\rho$, therefore it is continuous. For $\rho=0$ it is equal to 1 and as $\rho\to\infty$, it tends to $\infty$ as well. If we take its derivative with respect to $\rho$ at $\rho=0$ we obtain $\psi'(0)=\Exp_0[\tilde{\lambda}_T]$. For sufficiently large number of quantization levels $\tilde{\lambda}_T$ approximates $\lambda_T$ consequently $\psi'(0)\approx\Exp_0[\lambda_T]=T\Exp_0[\lambda_1]<0$ which is negative. This implies that, at least close to 0 and for positive values of $\rho$, the function $\psi(\rho)$ is decreasing and therefore strictly smaller than 1. Since we have values of $\rho$ for which $\psi(\rho)$ is smaller than 1 and other values for which it is larger than 1, due to continuity, there exists $\rho_0>0$ for which $\psi(\rho_0)=1$. In fact this $\rho_0>0$ is also unique due to convexity.

For any integer $m>0$ we have $\tilde{L}_{mT}=\sum_{n=1}^m\tilde{\lambda}_{nT}$ and, due to the definition of $\rho_0$ and the fact that $\{\tilde{\lambda}_{nT}\}$ is an i.i.d.~sequence we conclude that $\{e^{\rho_0\tilde{L}_{mT}}\}$ is a positive martingale which suggests that it is also a positive supermartingale. This allows us to apply the optional sampling theorem for positive submartingales which yields $\Exp_0[e^{\rho_0\tilde{L}_{\cM T}}]\leq1$ for any stopping time $\cM$ which is adapted to $\{\tilde{L}_{mT}\}$, as for instance the one in the definition of $\tilde{\cS}$ in \eqref{eq:Q-SPRT}. Because of this observation we can write
\ignore{
 and denote the pdf of these random variables under hypothesis $\Hyp_0$ as $f_0(\tilde{\lambda}_T,\ldots,\tilde{\lambda}_{mT})=\prod_{n=1}^m f_0(\tilde{\lambda}_{nT})$, with the latter equality being true since they are i.i.d. Using the fact $\tilde{L}_{mT}=\sum_{n=1}^m\tilde{\lambda}_{nT}$ we can then define a new pdf as follows
\begin{equation}
\tilde{f}_1(\tilde{\lambda}_T,\ldots,\tilde{\lambda}_{mT})
=e^{\rho_0\tilde{L}_{mT}}f_0(\tilde{\lambda}_T,\ldots,\tilde{\lambda}_{mT})
=\prod_{n=1}^m e^{\rho_0\tilde{\lambda}_{nT}}f_0(\tilde{\lambda}_{nT}).
\label{eq:appA.2}
\end{equation}
Indeed $\tilde{f}_1(\cdot)$ is a legitimate pdf since it is nonnegative and integrates to 1 because of the definition of $\rho_0$. Furthermore, under the new pdf, as we can see from \eqref{eq:appA.2}, the sequence $\{\tilde{\lambda}_{mT}\}$ is still i.i.d. This new pdf induces a new probability measure which we will denote as $\tilde{\Pro}_1$ and the corresponding expectation $\tilde{\Exp}_1[\cdot]$.
Applying change of measures and recalling that $\tilde{\cS}=\cM T$ we can write
}
\begin{equation}
\alpha=\Pro_0(\tilde{L}_{\tilde{\cS}}\ge\tilde{A})=\Pro_0({\rho_0\tilde{L}_{\tilde{\cS}}}\geq {\rho_0\tilde{A}})
\leq e^{-\rho_0\tilde{A}}\Exp_0[e^{\rho_0\tilde{L}_{\tilde{\cS}}}]=e^{-\rho_0\tilde{A}}\Exp_0[e^{\rho_0\tilde{L}_{\cM T}}]\leq e^{-\rho_0\tilde{A}},
\end{equation}
where for the first inequality we used the Markov inequality. Solving for $\tilde{A}$ yields \eqref{eq:appA.1}.

Let us now attempt to find a lower bound for $\rho_0$ as a function of the number $\tilde{r}$ of quantization levels.
We recall that $\rho_0$ is the solution of the equation $\Exp_0[e^{\rho\tilde{L}_T}]=1$. Note that $\Exp_0[e^{\rho\tilde{L}_T}]\leq\Exp_0[e^{\rho(L_T+TK\tilde{\epsilon})}]$, where $\tilde{\epsilon}=\frac{\phi}{\tilde{r}}$, consequently the positive solution $\tilde{\rho}$ of the equation $\psi(\rho)=\Exp_0[e^{\rho(L_T+TK\tilde{\epsilon})}]=1$ constitutes a lower bound for $\rho_0$, that is, $\rho_0\geq\tilde{\rho}_0$. The function $\psi(\rho)$ is convex and $\psi(1)>1$ which suggests $\tilde{\rho}_0<1$. Because of this observation we can write
\begin{equation}
1=\Exp_0[e^{\tilde{\rho}_0(L_T+TK\tilde{\epsilon})}]\leq\Exp_0[e^{\tilde{\rho}_0L_T+TK\tilde{\epsilon}}]=
\left(\Exp_0[e^{\tilde{\rho}_0L_1+K\tilde{\epsilon}}]\right)^T,
\label{eq:mbifla2}
\end{equation}
with the last equality being true because $L_T=l_1+\cdots+l_T$ with the $\{l_t\}$ being i.i.d.~thus suggesting $\Exp_0[e^{\tilde{\rho}_0L_T}]=(\Exp_0[e^{\tilde{\rho}_0l_1}])^T=(\Exp_0[e^{\tilde{\rho}_0L_1}])^T$ since $L_1=l_1$. From \eqref{eq:mbifla2} we can conclude
\begin{multline}
1\leq\Exp_0[e^{\tilde{\rho}_0L_1+K\tilde{\epsilon}}]=\Exp_1[e^{(\tilde{\rho}_0-1)L_1+K\tilde{\epsilon}}]
\leq\Exp_1\left[\frac{K\phi-L_1}{2K\phi}e^{-(\tilde{\rho}_0-1)K\phi+K\tilde{\epsilon}}+\frac{K\phi+L_1}{2K\phi}e^{(\tilde{\rho}_0-1)K\phi+K\tilde{\epsilon}}\right]\\
=\frac{\phi-\I_1}{2\phi}e^{-(\tilde{\rho}_0-1)K\phi+K\tilde{\epsilon}}+\frac{\phi+\I_1}{2\phi}e^{(\tilde{\rho}_0-1)K\phi+K\tilde{\epsilon}}.
\label{eq:mbifla}
\end{multline}
The second inequality comes from the convexity of the exponential function namely $\frac{a-z}{2a}e^{-a}+\frac{a+z}{2a}e^{a}\geq e^{-a\frac{a-z}{2a}+a\frac{a+z}{2a}}=e^z$, for $-a\leq z\leq a$. If we call $x=e^{(\tilde{\rho}_0-1)K\phi}$ then the inequality in \eqref{eq:mbifla} is equivalent to
\begin{equation}
\frac{\phi+\I_1}{2\phi}x^2-e^{-K\tilde{\epsilon}}x+\frac{\phi-\I_1}{2\phi}\geq0
\end{equation}
suggesting that $x$ is either larger than the largest root or smaller than the smallest root of the corresponding equation. We are interested in the first case namely
\begin{equation}
x=e^{(\tilde{\rho}_0-1)K\phi}\geq\frac{e^{-K\tilde{\epsilon}}+\sqrt{ e^{-2K\tilde{\epsilon}}-1+\frac{\I_1^2}{\phi^2} }}{1+\frac{\I_1}{\phi}}=1-K\left(\tilde{\epsilon}\frac{\phi}{\I_1}+o(\tilde{\epsilon})\right),
\end{equation}
where for the last equality we used the approximations $e^{\epsilon}=1+\epsilon+o(\epsilon)$ and $\sqrt{1+\epsilon}=1+0.5\epsilon+o(\epsilon)$. Taking now the logarithm, solving for $\tilde{\rho}_0$ and using the approximation $\log(1+\epsilon)=\epsilon+o(\epsilon)$ we end up with the lower bound $\tilde{\rho}_0\geq1-\frac{\tilde{\epsilon}}{\I_1}+o(\tilde{\epsilon})$. Finally recalling that $\rho_0\geq\tilde{\rho}_0$ proves the desired inequality.
\end{IEEEproof}

\begin{IEEEproof}[Proof of Theorem\,\ref{th:1}]
Again we will focus on the first inequality, the second can be shown similarly.
From the definition of $\tilde{\cS}$ in \eqref{eq:Q-SPRT} we have $\Exp_1[\tilde{\cS}]=T\Exp_1[\cM]$. From the definition of $\cM$ in \eqref{eq:Q-SPRT} and Wald's identity we can write
\begin{equation}
\Exp_1[\tilde{L}_{\cM T}]=\Exp_1\left[\sum_{m=1}^{\cM}\tilde{\lambda}_{mT}\right]=
\Exp_1[\cM]\Exp_1[\tilde{\lambda}_{T}],
\end{equation}
consequently
\begin{equation}
\Exp_1[\tilde{\cS}]=T\frac{\Exp_1[\tilde{L}_{\cM T}]}{\Exp_1[\tilde{\lambda}_{T}]}.
\label{eq:th.1.proof1}
\end{equation}

Next we upper bound the previous ratio.
Let us start with the denominator, for which we find the following lower bound
\begin{equation}
\Exp_1[\tilde{\lambda}_{T}]=\Exp_1[\lambda_{T}]+\Exp_1[\tilde{\lambda}_{T}-\lambda_{T}]\geq T\Exp_1[\lambda_1]-TK\tilde{\epsilon}=TK(\I_1-\tilde{\epsilon})=TK\I_1\left(1-\frac{\tilde{\epsilon}}{\I_1}\right),
\label{eq:th.1.proof2}
\end{equation}
where we recall $\tilde{\epsilon}=\frac{\phi}{\tilde{r}}$.

For the numerator we have the following upper bound
\begin{multline}
\Exp_1[\tilde{L}_{\cM T}]=\Exp_1[\tilde{L}_{\cM T}\ind{\tilde{L}_{\cM T}\geq\tilde{A}}]+\Exp_1[\tilde{L}_{\cM T}\ind{\tilde{L}_{\cM T}\leq-\tilde{B}}]\leq\Exp_1[\tilde{L}_{\cM T}\ind{\tilde{L}_{\cM T}\geq\tilde{A}}]\\
\leq\tilde{A}+\Exp_1[(\tilde{L}_{\cM T}-\tilde{A})\ind{\tilde{L}_{\cM T}\geq\tilde{A}}]
=\tilde{A}+\Exp_1[(\tilde{L}_{(\cM-1)T}-\tilde{A}+\tilde{\lambda}_{\cM T})\ind{\tilde{L}_{\cM T}\geq\tilde{A}}].
\label{eq:th1.2}
\end{multline}
The second term in the right hand side of the first equality is negative, therefore, eliminating it yields an upper bound. Note now that since $\cM T$ is the first time $\tilde{L}_{\cM T}$ exceeds $\tilde{A}$ we necessarily have
$\tilde{L}_{(\cM-1)T}-\tilde{A}\leq0$. Also $\tilde{\lambda}_{mT}=\sum_{k=1}^K\tilde{\lambda}^k_{mT}\leq KT\phi$ since, as we have seen, $\tilde{\lambda}^k_{mT}\leq T\phi$. These two observations combined with Lemma\,\ref{lem:appA.1} and used in \eqref{eq:th1.2} suggest
\begin{equation}
\Exp_1[\tilde{L}_{\cM T}]\leq\tilde{A}+KT\phi\leq \frac{|\log\alpha|}{1-\frac{\tilde{\epsilon}}{\I_1}\{1+o_{\tilde{r}}(1)\}}+KT\phi=|\log\alpha|\left\{1+\frac{\tilde{\epsilon}}{\I_1}\{1+o_{\tilde{r}}(1)\}\right\}+KT\phi.
\end{equation}
Applying in \eqref{eq:th.1.proof1} the previous bound and the bound in \eqref{eq:th.1.proof2}, yields
\begin{equation}
\Exp_1[\tilde{\cS}]\le\frac{|\log\alpha|}{K\I_1}\frac{1+\frac{\tilde{\epsilon}}{\I_1}\{1+o_{\tilde{r}}(1)\}}{1-\frac{\tilde{\epsilon}}{\I_1}}+T\frac{\phi}{\I_1}\frac{1}{1-\frac{\tilde{\epsilon}}{\I_1}}
=\frac{|\log\alpha|}{K\I_1}\left\{1+2\frac{\tilde{\epsilon}}{\I_1}\{1+o_{\tilde{r}}(1)\}\right\}+T\frac{\phi}{\I_1}\{1+o_{\tilde{r}}(1)\}.
\end{equation}
Finally using \eqref{eq:SPRTperf_d} from Lemma\,\ref{lem:SPRT_d}, we obtain
\begin{equation}
0\leq\Exp_1[\tilde{\cS}]-\Exp_1[\cS]\leq\frac{2|\log\alpha|}{K\I_1^2}\frac{\phi}{\tilde{r}}\{1+o_{\tilde{r}}(1)\}+T\frac{\phi}{\I_1}\{1+o_{\tilde{r}}(1)\}+o_{\alpha}(1),
\end{equation}
which is the desired inequality.
\end{IEEEproof}

\section*{\vskip-1cm Appendix B}
Before proving Theorem\,\ref{th:2} we need to present a number of technical lemmas.

\begin{lem}\label{lem:appB.1}\vskip-0.3cm
Consider the sequence $\{\hat{L}_{t_n}\}$ defined in \eqref{eq:LT-update} where $\{t_n\}$ is the increasing sequence of time instants at which the FC receives information from some SU. We then have that $\{e^{\hat{L}_{t_n}}\}$ and $\{e^{-\hat{L}_{t_n}}\}$ are supermartingales in $n$ with respect to the probability measures $\Pro_0$ and $\Pro_1$ respectively where the two measures also account for the randomizations.
\end{lem}
\begin{IEEEproof}
We will show the first claim namely $\Exp_0[e^{\hat{L}_{t_n}}|\hat{L}_{t_{n-1}},\ldots,\hat{L}_{t_{1}}]\leq e^{\hat{L}_{t_{n-1}}}$. It is sufficient to prove that $\Exp_0[e^{\hat{L}_{t_n}-\hat{L}_{t_{n-1}}}|\hat{L}_{t_{n-1}},\ldots,\hat{L}_{t_{1}}]\leq 1$ and, using \eqref{eq:LT-update}, that
$\Exp_0[e^{b_n(\Delta+\hat{q}_n)}|\hat{L}_{t_{n-1}},\ldots,\hat{L}_{t_{1}}]\leq 1$.

Let $\{(b_1,\hat{q}_1),\ldots,(b_n,\hat{q}_n)\}$ denote the $n$ messages received by the fusion center until the $n$th communication time $t_n$. Denote with $\{k_1,\ldots,k_n\}$ the indices of the corresponding transmitting SUs. Then the $n$ messages \textit{given} these indices are \textit{independent} due to the independence of observations across time and across SUs. Using the tower property of expectation we can then write
\begin{multline}
\Exp_0[e^{b_n(\Delta+\hat{q}_n)}|\hat{L}_{t_{n-1}},\ldots,\hat{L}_{t_{1}}]\\
=\Exp_0[\Exp_0[e^{b^{k_n}_n(\Delta+\hat{q}^{k_n}_n)}|(k_n,b_n^{k_n},q_n^{k_n},\hat{q}_n^{k_n}),\ldots,(k_1,b_1^{k_1},q_1^{k_1},\hat{q}_1^{k_1})]|\hat{L}_{t_{n-1}},\ldots,\hat{L}_{t_{1}}]\\
=\Exp_0[\Exp_0[e^{b^{k_n}_n(\Delta+\hat{q}^{k_n}_n)}|k_n,b_n^{k_n},q_n^{k_n},\hat{q}_n^{k_n}]]
=\Exp_0[\Exp_0[e^{b^{k_n}_n(\Delta+\hat{q}^{k_n}_n)}|k_n,b_n^{k_n},q_n^{k_n}]]\\
=\Exp_0[\Exp_0[\Exp[e^{b^{k_n}_n(\Delta+\hat{q}^{k_n}_n)}|b_n^{k_n},q_n^{k_n}]|k_n]]
\leq\Exp_0[\Exp_0[e^{L^{k_n}_{t_n^{k_n}}-L^{k_n}_{t_{n-1}^{k_n}}}|k_n]],
\end{multline}
with the second equality being true due to the conditional independence of the messages; the last inequality due to \eqref{eq:lem1.1} of Lemma\,\ref{lem:quant}; and $\Exp[\cdot]$ denoting expectation with respect to the randomization.
Now $\{e^{L_t^k}\}$ is a martingale with respect to $\Pro_0$, therefore it is also a supermartingale. For stopping times  $\cT_1>\cT_2$ we have from optional sampling for positive supermartingales
$\Exp_0[e^{L^{k_n}_{\cT_1}}|k_n,L^{k_n}_1,\ldots,L^{k_n}_{\cT_2}]\leq e^{L^{k_n}_{\cT_2}}$ from which we conclude that
$\Exp_0[e^{L^{k_n}_{\cT_1}-L^{k_n}_{\cT_2}}|k_n]\leq1$. Our lemma is proven by selecting $\cT_1=t_n^{k_n}$ and $\cT_2=t_{n-1}^{k_n}$.
\end{IEEEproof}

An immediate consequence of the previous lemma and the application of optional sampling for positive supermartingales is the following corollary.

\begin{cor}\label{cor:1}\vskip-0.3cm
If $\cN\ge0$ is any stopping time which depends on the process $\{\hat{L}_{t_n}\}$ and since $t_0=0$ and $\hat{L}_0=0$, we conclude that
\begin{equation}
\Exp_0[e^{\hat{L}_{t_{\cN}}}]\leq e^{\hat{L}_{t_0}}=1;~~\Exp_1[e^{-\hat{L}_{t_{\cN}}}]\leq e^{-\hat{L}_{t_0}}=1.
\end{equation}
In particular for the case $\cN=\inf\{n:\hat{L}_{t_n}\not\in(-\hat{B},\hat{A})\}$ and recalling the definition of the RLT-SPRT stopping time $\hat{\cS}=t_{\cN}$ in \eqref{eq:RLT-SPRT}, we have
\begin{equation}
\Exp_0[e^{\hat{L}_{\hat{\cS}}}]\le1;~~\Exp_1[e^{-\hat{L}_{\hat{\cS}}}]\le1.
\label{eq:cor1.1}
\end{equation}
\end{cor}
Let us now find useful estimates for RLT-SPRT.
\begin{lem}\label{lem:dontknow}\vskip-0.3cm
If $\hat{A},\hat{B}$ are selected in RLT-SPRT to assure error probabilities $\alpha,\beta$ then
\begin{gather}
\hat{A}\leq |\log\alpha|;~~~
\hat{B}\leq |\log\beta|,
\label{eq:dontknow1}\\
\Exp_1[\hat{L}_{\hat{\cS}}]\leq|\log\alpha|+K\phi;~~~
-\Exp_0[\hat{L}_{\hat{\cS}}]\leq|\log\beta|+K\phi.
\label{eq:dontknow2}
\end{gather}
\end{lem}
\begin{IEEEproof}
According to our usual practice we will only show the first inequality in both cases. For \eqref{eq:dontknow1} note that
\begin{equation}
\alpha=\Pro_0(\hat{L}_{\hat{\cS}}\geq\hat{A})\leq e^{-\hat{A}}\Exp_0[e^{\hat{L}_{\hat{\cS}}}]\leq e^{-\hat{A}},
\end{equation}
where we used the Markov inequality and \eqref{eq:cor1.1}.

For \eqref{eq:dontknow2} we can write
\begin{align}
\Exp_1[\hat{L}_{\hat{\cS}}]&=\Exp_1[\hat{L}_{\hat{\cS}}\ind{\hat{L}_{\hat{\cS}}\geq\hat{A}}]+
\Exp_1[\hat{L}_{\hat{\cS}}\ind{\hat{L}_{\hat{\cS}}\leq-\hat{B}}]\leq\Exp_1[\hat{L}_{\hat{\cS}}\ind{\hat{L}_{\hat{\cS}}\geq\hat{A}}] \leq \hat{A}+\Exp_1[(\hat{L}_{\hat{\cS}}-\hat{A})\ind{\hat{L}_{\hat{\cS}}\geq\hat{A}}]\nn\\
&\leq\hat{A}+K\phi\leq|\log\alpha|+K\phi.
\label{eq:dontknow3}
\end{align}
The first inequality in \eqref{eq:dontknow3} comes from the fact that the overshoot cannot exceed the last update performed by the FC on its test statistic $\{\hat{L}_t\}$. The maximum value of this update is $K\phi$ since we can have, at most, all SUs transmitting information to the FC and each message is upper bounded by $\phi$. Of course for the last inequality we used \eqref{eq:dontknow1}.
\end{IEEEproof}

\begin{lem}\label{lem:wald}\vskip-0.3cm
Let $\{t_n^k\}$ be the sequence of sampling times at the $k$th SU and denote with $\cN^k_t$ the number of samples taken up to time $t$. Consider an i.i.d. sequence of random variables $\{\zeta_n\}$ where each $\zeta_n$ is a bounded function of the observations $y^k_{t^k_{n-1}+1},\ldots,y^k_{t^k_{n}}$ such that $|\zeta_n|\leq M<\infty$. Let $\cT$ be a stopping time which at every time instant $t$ depends on the global information from all SUs up to time $t$. Then we have the following version of Wald's identity
\begin{gather}
\Exp_i\left[\sum_{n=1}^{\cN^k_{\cT}}\zeta_n\right]\geq\Exp_i[\zeta_1]\Exp_i[\cN^k_{\cT}]-2M.
\label{eq:wald}
\end{gather}
\end{lem}
\begin{IEEEproof}
The proof can be found in \cite[Lemma\,3]{Fellouris11}.
\end{IEEEproof}

Next we estimate the average sampling period of each SU.

\begin{lem}\label{lem:period}\vskip-0.3cm
Let $\{t_n^k\}$ be the sequence of sampling times at the $k$th SU, with $\Delta$ the common parameter that defines the local thresholds, then
\begin{align}
|\Exp_i[L^k_{t^k_n}-L^k_{t^k_{n-1}}]|&=|\Exp_i[L^k_1]|\Exp_i[t^k_n-t^k_{n-1}]\geq\Delta\tanh\left(\frac{\Delta}{2}\right),\label{eq:lem.period1}\\
|\Exp_i[\hat{L}^k_{t^k_n}-\hat{L}^k_{t^k_{n-1}}]|&\geq\Delta\tanh\left(\frac{\Delta}{2}\right)-\hat{\epsilon}=\Delta\tanh\left(\frac{\Delta}{2}\right)\{1+o_{\Delta,\hat{r}}(1)\}.
\label{eq:lem.period2}
\end{align}
\end{lem}
\begin{IEEEproof}
As we mentioned before, the sampling process at each SU is based on a repeated SPRT with thresholds $-\Delta,\Delta$, and every time this SPRT stops we sample the incremental LLR process $L^k_t-L^k_{t^k_{n-1}}$. Using the classical Wald identity and the known lower bounds for the corresponding average delays we have under $\Hyp_1$
\begin{equation}
\Exp_1[L^k_{t^k_n}-L^k_{t^k_{n-1}}]=\Exp_1[L^k_1]\Exp_1[t^k_n-t^k_{n-1}]\geq\cH(\beta_k,\alpha_k)
\label{eq:appB.3}
\end{equation}
where, we recall, $\cH(x,y)=x\log \frac{x}{1-y}+(1-x)\log\frac{1-x}{y}$ and $\alpha_k=\Pro_0(L^k_{t^k_n}-L^k_{t^k_{n-1}}\geq\Delta)$, $\beta_k=\Pro_1(L^k_{t^k_n}-L^k_{t^k_{n-1}}\leq-\Delta)$.

From Wald's classical estimate of the error probabilities we have $\alpha_k\leq e^{-\Delta}(1-\beta_k)$ and $\beta_k\leq e^{-\Delta}(1-\alpha_k)$. These two inequalities generate the following two regions of points: 1)~for $0\leq\beta_k\leq \frac{1}{1+e^{\Delta}}$ we have $0\le\alpha_k\leq e^{-\Delta}(1-\beta_k)$; 2)~for $\frac{1}{1+e^{\Delta}}\leq\beta_k\leq e^{-\Delta}$ we have $0\leq\alpha_k\leq 1-\beta_k e^{\Delta}$. Since we cannot compute the exact values for the two error probabilities $\alpha_k,\beta_k$, we will find the worst possible pair $(\alpha_k,\beta_k)$ within the two regions that minimizes the lower bound $\cH(\beta_k,\alpha_k)$.

The function $\cH(\beta_k,\alpha_k)$ is decreasing in both its arguments, provided that $\alpha_k+\beta_k\leq1$. Therefore when $0\leq\beta_k\leq\frac{1}{1+e^{\Delta}}$ we can replace $\alpha_k$ with its maximal value $e^{-\Delta}(1-\beta_k)$ and strengthen the inequality in \eqref{eq:appB.3}. The resulting lower bound $\cH(\beta_k,e^{-\Delta}(1-\beta_k))$ as a function of $\beta_k$ is decreasing and therefore exhibits its minimum for $\beta_k=\frac{1}{1+e^{\Delta}}$. Similarly when $\frac{1}{1+e^{\Delta}}\leq\beta_k\leq e^{-\Delta}$ we can replace again $\alpha_k$ with its maximal value and strengthen the inequality. The corresponding lower bound is now $\cH(\beta_k,1-\beta_k e^{\Delta})$ which, as a function of $\beta_k$ is increasing, therefore the minimum appears again for $\beta_k=\frac{1}{1+e^{\Delta}}$. This suggests that the lower bound is minimized when $\beta_k=\frac{1}{1+e^{\Delta}}$ which, in both cases yields an equal value for $\alpha_k$ namely $\alpha_k=\frac{1}{1+e^{\Delta}}$. Concluding, the final lower bound is $\cH(\frac{1}{1+e^{\Delta}},\frac{1}{1+e^{\Delta}})$, which is equal to $\Delta\tanh(\frac{\Delta}{2})$. Similarly we can show the bound under $\Hyp_0$.

Proving \eqref{eq:lem.period2} is straightforward since the difference $\hat{L}^k_{t^k_n}-\hat{L}^k_{t^k_{n-1}}$ is simply the quantized version of $L^k_{t^k_n}-L^k_{t^k_{n-1}}$ and, by design, the quantization error does not exceed $\hat{\epsilon}=\frac{\phi}{\hat{r}}$.
\end{IEEEproof}

\begin{IEEEproof}[Proof of Theorem\,\ref{th:2}]
We need to find an upper bound for $\Exp_1[\hat{\cS}]$. Note that using the classical Wald identity we can write
\begin{equation}
K\I_1\Exp_1[\hat{\cS}]=\Exp_1[L_{\hat{\cS}}]=\Exp_1[L_{\hat{\cS}}-\hat{L}_{\hat{\cS}}]+\Exp_1[\hat{L}_{\hat{\cS}}].
\label{eq:prTH2.1}
\end{equation}
Let us consider the term $L_{t}-\hat{L}_{t}=\sum_{k=1}^K(L^k_{t}-\hat{L}^k_{t})$. For the $k$th SU we have the sequence of sampling times $\{t^k_n\}$; call $\cN_t^k$ the number of samples taken up to (and including) time $t$. Then we can write
\begin{equation}
L^k_{t}-\hat{L}^k_{t}=(L^k_t-L^k_{t_{\cN^k_t}})+(L^k_{t^k_{\cN^k_t}}-\hat{L}^k_{t^k_{\cN^k_t}}),
\end{equation}
with the equality being true because $\hat{L}^k_t=\hat{L}^k_{t^k_{\cN^k_t}}$. The first term in the right hand side is the incremental LLR at the $k$th SU before the next sampling. Since this quantity lies in the interval $(-\Delta,\Delta)$ it is upper bounded by $\Delta$. Consequently we can write
\begin{equation}
L^k_{t}-\hat{L}^k_{t}\leq\Delta+\sum_{n=1}^{\cN^k_t}\{L^k_{t^k_{n}}-L^k_{t^k_{n-1}}-(\hat{L}^k_{t^k_{n}}-\hat{L}^k_{t^k_{n-1}})\}\leq\Delta+\sum_{n=1}^{\cN^k_t}\{b_n^k(\Delta+q_n^k)-b_n^k(\Delta+\hat{q}_n^k)\}\leq\Delta+\hat{\epsilon}\cN^k_t,
\end{equation}
where we recall that $\hat{\epsilon}=\frac{\phi}{\hat{r}}$ is the maximal quantization error. Replacing $t$ with $\hat{\cS}$, taking expectation on both sides and summing over $k$ yields
\begin{equation}
\Exp_1[L_{\hat{\cS}}-\hat{L}_{\hat{\cS}}]\leq K\Delta+\hat{\epsilon}\Exp_1[\cN_{\hat{\cS}}].
\label{eq:prTH2.2}
\end{equation}
where $\cN_{\hat{\cS}}$ is the total number of messages received by the FC up to the time of stopping.

Consider now the following expectation and use \eqref{eq:wald} from Lemma\,\ref{lem:wald} and \eqref{eq:lem.period2} from Lemma\,\ref{lem:period}
\begin{align}
\Exp_1[\hat{L}^k_{\hat{\cS}}]&=\Exp_1[\hat{L}^k_{t^k_{\cN^k_{\hat{\cS}}}}]=\Exp_1\left[\sum_{n=1}^{\cN^k_{\hat{\cS}}}(\hat{L}^k_{t^k_{n}}-\hat{L}^k_{t^k_{n-1}})\right]
\geq\Exp_1[\hat{L}^k_{t^k_{1}}]\Exp_1[\cN^k_{\hat{\cS}}]-2\phi\nn\\
&\geq \left(\Delta\tanh\left(\frac{\Delta}{2}\right)-\hat{\epsilon}\right)\Exp_1[\cN^k_{\hat{\cS}}]-2\phi.
\end{align}
Summing over $k$ and solving for $\Exp_1[\cN_{\hat{\cS}}]$ yields
\begin{equation}
\Exp_1[\cN_{\hat{\cS}}]\leq \frac{\Exp_1[\hat{L}_{\hat{\cS}}]+2K\phi}{\Delta\tanh(\frac{\Delta}{2})-\hat{\epsilon}}.
\end{equation}
Replacing this in \eqref{eq:prTH2.2}, we obtain
\begin{equation}
\Exp_1[L_{\hat{\cS}}-\hat{L}_{\hat{\cS}}]\leq K\Delta+\hat{\epsilon}\frac{\Exp_1[\hat{L}_{\hat{\cS}}]+2K\phi}{\Delta\tanh(\frac{\Delta}{2})-\hat{\epsilon}}.
\end{equation}
Finally using the previous inequality in \eqref{eq:prTH2.1} and \eqref{eq:dontknow2} from Lemma\,\ref{lem:dontknow}, yields
\begin{multline}
K\I_1\Exp_1[\hat{\cS}]\leq \left(1+\frac{\hat{\epsilon}}{\Delta\tanh(\frac{\Delta}{2})-\hat{\epsilon}}\right)|\log\alpha|+K\left(\Delta+\phi+\frac{3\phi\hat{\epsilon}}{\Delta\tanh(\frac{\Delta}{2})-\hat{\epsilon}}\right)\\
=\left(1+\frac{\hat{\epsilon}\{1+o_{\Delta,\hat{r}}(1)\}}{\Delta\tanh(\frac{\Delta}{2})}\right)|\log\alpha|+K(\Delta+\phi+o_{\Delta,\hat{r}}(1))
\end{multline}
Subtracting the lower bound \eqref{eq:SPRTperf_d} for the optimum $\Exp_1[\cS]$ we obtain the desired estimate.
\end{IEEEproof}


\end{document}